\documentclass[a4paper,fleqn,usenatbib,useAMS]{mnras}
\usepackage{graphicx}	
\usepackage{amsmath}	
\usepackage{amssymb}	
\usepackage{multicol}        
\usepackage{bm}		
\usepackage{pdflscape}	
\usepackage[T1]{fontenc}
\usepackage{ae,aecompl}
\usepackage{txfonts}
\usepackage{threeparttable}
\usepackage{makecell}

\title[BL~Her models]
{A theoretical framework of BL~Her stars. I. Effect of metallicity and convection parameters on period-luminosity and period-radius relations}
\author[Das et al.]{Susmita Das$^{1}$\thanks{E-mail: susmitadas130\@gmail.com}, Shashi M. Kanbur$^{2}$, Radoslaw Smolec$^{3}$, Anupam Bhardwaj$^{4}$, 
\newauthor
Harinder P. Singh$^{1}$ and Marina Rejkuba$^{5}$\\
\vspace{1pt}\\
1. Department of Physics \& Astrophysics, University of Delhi, Delhi 110007, India \\
2. Department of Physics, State University of New York Oswego, Oswego, NY 13126, USA \\
3. Nicolaus Copernicus Astronomical Center of the Polish Academy of Sciences, Bartycka 18, PL-00-716 Warszawa, Poland \\
4. Kavli Institute for Astronomy and Astrophysics, Peking University, Yi He Yuan Lu 5, Hai Dian District, Beijing 100871, China\\
5. European Southern Observatory, Karl-Schwarzschild-Stra{\ss}e 2, D-85748 Garching bei M{\"u}nchen, Germany
}

\begin{document}

\date{Accepted 2020 November 23. Received 2020 November 19; in original form 2020 September 10}

\pagerange{\pageref{firstpage}--\pageref{lastpage}} \pubyear{2020}

\maketitle

\label{firstpage}

\begin{abstract}
We present a new grid of convective BL~Herculis models using the state-of-the-art 1D non-linear radial stellar pulsation tool \textsc{mesa-rsp}. We investigate the impact of metallicity and four sets of different convection parameters on multi-wavelength properties. Non-linear models were computed for periods typical for BL~Her stars, i.e. $1 \leq \mathrm{P (days)} \leq 4$ covering a wide range of input parameters - metallicity ($-2.0\; \mathrm{dex} \leq \mathrm{[Fe/H]} \leq 0.0\; \mathrm{dex}$), stellar mass (0.5M$_{\odot}$-0.8M$_{\odot}$), luminosity (50L$_{\odot}$-300L$_{\odot}$) and effective temperature (full extent of the instability strip; in steps of 50K). The total number of BL~Her models with full-amplitude stable pulsations used in this study is 10280 across the four sets of convection parameters. We obtain their multiband ($UBVRIJHKLL'M$) light curves and derive new theoretical period-luminosity ($PL$), period-Wesenheit ($PW$) and period-radius ($PR$) relations at mean light. We find that the models computed with radiative cooling show statistically similar slopes for $PL$, $PW$ and $PR$ relations. Most empirical relations match well with the theoretical $PL$, $PW$ and $PR$ relations from the BL~Her models computed using the four sets of convection parameters. However, $PL$ slopes of the models with radiative cooling provide a better match to empirical relations for BL~Her stars in the LMC in the $HK_S$ bands. For each set of convection parameters, the effect of metallicity is significant in $U$ and $B$-bands and negligible in infrared bands, which is consistent with empirical results. No significant metallicity effects are seen in the $PR$ relations.
\end{abstract}


\section{Introduction}

Type~II Cepheids (T2Cs) are pulsating stars located in the instability strip of the Hertzsprung-Russell diagram (HRD). T2Cs are brighter than the RR~Lyrae stars but fainter than the classical Cepheids. Based on their pulsational period, T2Cs are divided into the following subclasses: the BL~Herculis (BL~Her: $1 \lesssim P \textrm{(days)}\lesssim 4$), the W~Virginis (W~Vir: $4 \lesssim P \textrm{(days)}\lesssim 20$) and the RV~Tauris (RV~Tau: $P \gtrsim 20$ days) \citep{soszynski2018}. However, the period separation for different subclasses is not strict, for example, the upper limit on the period of BL Her stars is set at 4 days in the Magellanic Clouds \citet{soszynski2018} and 5 days in the Galactic bulge \citet{soszynski2017}. Similar to RR~Lyrae stars and classical Cepheids, T2Cs follow well-defined period-luminosity ($PL$) relationships \citep{matsunaga2006, groenewegen2008, matsunaga2009, ciechanowska2010, matsunaga2011, ripepi2015, bhardwaj2017b, bhardwaj2017c, groenewegen2017b, braga2018}, which makes them useful distance indicators \citep[see reviews,][]{beaton2018b, bhardwaj2020a}.

T2Cs are population II stars and trace low mass, metal-poor, old-age stellar populations. However, recent studies suggest that W~Vir stars may have their origin in binary systems \citep{groenewegen2017a} and that RV~Tau stars may also have massive and younger progenitors \citep{manick2018}. T2Cs have a wide range of metallicities \citep{welch2012}. While \citet{clement2001} noted that all the Galactic Globular Clusters (GGCs) containing T2Cs have $\mathrm{[Fe/H]}<-1.0$ dex, Galactic field T2Cs were found to have metallicities in the range $-1.0\; \mathrm{dex}<\mathrm{[Fe/H]}<0\; \mathrm{dex}$ \citep{schmidt2011}. T2Cs in the Bulge have photometric metallicities ranging from $-1.4$ dex to $+0.6$ dex, with $\mathrm{[Fe/H]_{mean}} = -0.6 \pm 0.17$ dex \citep{harris1984, wallerstein2002}. In a recent study to understand the origin of the Galactic Halo, \citet{wallerstein2018} found $\mathrm{[Fe/H]}>-0.9$ dex for a majority of the field T2Cs. We therefore adopt a broad range of metallicities, $-2\; \mathrm{dex} <\mathrm{[Fe/H]}<0\; \mathrm{dex}$ for computing BL~Her models in the present study.

Empirical $PL$ relations of T2Cs have been studied extensively in the last few decades. \citet{nemec1994} provided $P-L-\mathrm{[Fe/H]}$ relations for T2Cs in the GGCs in the optical $BV$ bands. \citet{matsunaga2006} did not find any significant metallicity effect on the $PL$ relations in their study of $PL$ relations using 46 T2Cs in 26 GGCs; the observations were obtained from the Infrared Survey Facility (IRSF) 1.4-m telescope in the near-infrared $JHK_S$ bands. Using 39 T2Cs in the Galactic Bulge monitored with the SOFI infrared camera on the 3.5-m NTT on ESO/La Silla, \citet{groenewegen2008} provided $PL$ relations in the $K$-band and estimated a distance modulus of $14.51 \pm 0.12 \pm 0.07$ (systematic) mag to the Galactic Centre. The $PL$ relations for Galactic T2Cs in the Gaia $G$-band have been provided by \citet{clementini2016}. In a series of papers, \citet{matsunaga2009,matsunaga2011} presented near-infrared ($JHK_S$) $PL$ and Wesenheit relations for T2Cs in the Large (LMC) and Small (SMC) Magellanic Clouds obtained using data from IRSF and SIRIUS. Near-infrared $PL$ and period-Wesenheit ($PW$) relations for T2Cs in the LMC have been presented by \citet{ripepi2015} using the VISTA Magellanic Cloud survey \citep[VMC,][]{cioni2011} and by \citet{bhardwaj2017b} using observations obtained by the Large Magellanic Cloud Near-infrared Synoptic Survey \citep[LMCNISS,][]{macri2015}. Recently, \citet{manick2017} published $PW$ relations for T2Cs in the LMC using the Optical Gravitational Lensing Experiment \citep[OGLE-III,][]{soszynski2008} data, while \citet{groenewegen2017b} presented $PL$ relations of the Magellanic Cloud T2Cs based on OGLE-III data reporting no dependence on metallicity.

\citet{burki1986} and \citet{balog1997} are few of the earlier studies where the empirical period-radius ($PR$) relations of T2Cs were investigated. A detailed study of the $PR$ relations of T2Cs in the Magellanic Clouds has been carried out by \cite{groenewegen2017b} based on OGLE-III data. They found the $PR$ relations to have little or no dependence on metallicity.

On the theoretical front, several linear and non-linear convective T2C models, in particular, BL~Her models have been computed by \citet{bono1995, bono1997a, bono1997b, marconi2007, smolec2012a, smolec2012b, smolec2014, smolec2016}. \citet{buchler1992} had predicted period doubling in BL~Her stars which is caused by the 3:2 resonance between the fundamental mode and the first overtone. This was confirmed almost 20 years later when the period doubling behaviour was observed in a 2.4-d BL~Her type variable in the Galactic bulge and consistently modeled with the observed light curves \citep{soszynski2011, smolec2012a}. Few theoretical studies have provided $PL$ and $PR$ relations for BL~Her models. Theoretical near-infrared period-magnitude and $PW$ relations for BL~Her models in the metal abundance range of $Z=0.0001$ to $Z=0.004$ \footnote{Equivalent metallicity range, $-2.62\; \mathrm{dex} \leq \mathrm{[Fe/H]} \leq -0.66\; \mathrm{dex}$} were derived by \citet{criscienzo2007} while \citet{marconi2007} presented theoretical $PR$ relation for BL~Her models and found it to be in excellent agreement with the empirical relation from \citet{burki1986}.

The recently released non-linear Radial Stellar Pulsation (\textsc{rsp}) tool in \emph{Modules for Experiments in Stellar Astrophysics} \citep[\textsc{mesa},][]{paxton2011,paxton2013,paxton2015,paxton2018,paxton2019} may be used for generating multi-wavelength light curves of classical pulsators. Along with being an open-source code, \textsc{mesa-rsp} offers the advantage of testing how properties of the models depend on the details of convection model used, by varying convective parameters. \citet{das2020} computed a few RR~Lyrae, BL~Her and classical Cepheid models using \textsc{mesa-rsp} and found the theoretical period-colour (PC) relations to be in good agreement with the empirical PC relations. The aim of this project is to compute a very fine grid of BL~Her models, encompassing a wide range of metallicity, mass, luminosity and effective temperature using the most recent, state-of-the-art stellar pulsation code, the \textsc{mesa-rsp}. We also obtain theoretical $PL$ and $PR$ relations for these stars and test the effect of convection parameters and metallicity on these relations. The reason for choosing BL~Her stars only (and not the other subclasses of T2Cs) for our study is two-fold: (i) \citet{matsunaga2011} found evidence that $PL$ relations of BL~Her and W~Vir stars should be discussed independently (ii) The highly non-adiabatic longer-period T2Cs (W~Vir and RV~Tau stars) pose problems for the existing pulsation codes and are current limitations of \textsc{mesa-rsp} \citep{smolec2016,paxton2019}. However, \textsc{mesa-rsp} may be reliably used for modelling the shortest-period class of T2Cs, the BL~Her stars.

The structure of this paper is as follows: The BL~Her models computed using \textsc{mesa-rsp} are described in Section~\ref{sec:data}. In Sections~\ref{sec:PL} and \ref{sec:PR}, we study the $PL$ and $PR$ relations of these models and investigate any possible dependence of these relations on metallicity and convection parameters. Finally, we summarise the results of this study in Section~\ref{sec:results}.

\section{The stellar pulsation models}
\label{sec:data}

Since \textsc{mesa-rsp} offers the possibility of using different convection parameter sets, we explore the effect of different convection parameters on the multi-wavelength $PL$ and $PR$ relations of a finely computed grid of BL~Her models. We note here that \textsc{mesa-rsp} uses the theory of turbulent convection as outlined in \citet{kuhfuss1986} and follows \citet{smolec2008} in its treatment of stellar pulsation. The free parameters that enter the convective model are provided in Tables~3 and 4 of \citet{paxton2019}. For convenience, they are listed in Table~\ref{tab:convection}. Set A corresponds to the simplest convection model, set B adds radiative cooling, set C adds turbulent pressure and turbulent flux, and set D includes these effects simultaneously. A detailed description of the free parameters and their standard values is provided in \citet{smolec2008}. In brief, parameters $\alpha_p$ and $\alpha_c$ were introduced by \citet{yecko1998} and their values were set at $\alpha_p$=2/3 and $\alpha_c$=$\alpha_s$. The value for $\gamma_r$=$2\sqrt{3}$ is obtained from \citet{wuchterl1998}. Neglecting radiative cooling and turbulent pressure reduces the time-independent version of the convection model \citep{kuhfuss1986} to the standard mixing-length theory (MLT), provided the values for $\alpha_s$, $\alpha_c$ and $\alpha_d$ are kept the same as in Table~\ref{tab:convection}. \citet{paxton2019} suggest $\alpha_t \simeq 0.01$, $\alpha_m \lesssim 1$, and $\alpha \lesssim 2$ as useful starting choices. We stress here that we have not made any changes to the free parameters in this work and have used the four sets of convection parameters as provided in \citet{paxton2019}. We use \textsc{mesa}~r$11701$ for our present study. OPAL opacity tables \citep{iglesias1996} supplemented at low temperatures with \citet{ferguson2005} opacity data were adopted.

\begin{table}
\caption{The free parameters and their associated values in the convective parameter sets A, B, C and D of the \textsc{mesa-rsp} convection model.}
\centering
\scalebox{0.85}{
\begin{tabular}{l l c c c c}
\hline\hline
Name & Parameter description & Set A & Set B & Set C & Set D\\
\hline \hline
$\alpha$ & Mixing-length parameter & 1.5 & 1.5 & 1.5 & 1.5 \\
$\alpha_m$ & Eddy-viscous dissipation parameter & 0.25 & 0.50 & 0.40 & 0.70 \\
$\alpha_s$ & Turbulent source parameter & $\frac{1}{2}\sqrt{\frac{2}{3}}$ & $\frac{1}{2}\sqrt{\frac{2}{3}}$ & $\frac{1}{2}\sqrt{\frac{2}{3}}$ & $\frac{1}{2}\sqrt{\frac{2}{3}}$\\
$\alpha_c$ & Convective flux parameter & $\frac{1}{2}\sqrt{\frac{2}{3}}$ & $\frac{1}{2}\sqrt{\frac{2}{3}}$ & $\frac{1}{2}\sqrt{\frac{2}{3}}$ & $\frac{1}{2}\sqrt{\frac{2}{3}}$\\
$\alpha_d$ & Turbulent dissipation parameter & $\frac{8}{3}\sqrt{\frac{2}{3}}$ & $\frac{8}{3}\sqrt{\frac{2}{3}}$ & $\frac{8}{3}\sqrt{\frac{2}{3}}$ & $\frac{8}{3}\sqrt{\frac{2}{3}}$\\
$\alpha_p$ & Turbulent pressure parameter & 0 & 0 & $\frac{2}{3}$ & $\frac{2}{3}$ \\
$\alpha_t$ & Turbulent flux parameter & 0 & 0 & 0.01 & 0.01 \\
$\gamma_r$ & Radiative cooling parameter & 0 & $2\sqrt{3}$ & 0 & $2\sqrt{3}$\\
\hline
\end{tabular}}
\label{tab:convection}
\end{table}

\subsection{Linear computations}

We compute a fine grid of BL~Her models for each of the four convection sets with the following input parameters:

\begin{enumerate}
\item Metallicity (Corresponding $ZX$ values are provided in Table~\ref{tab:composition})\\

$\mathrm{[Fe/H]\; \mathrm{(in\; dex)}}= -2, -1.5, -1.35, -1, -0.5, -0.2, 0$ \\

\begin{table}
\caption{Chemical compositions of the adopted pulsation models$^*$.}
\centering
\begin{tabular}{c c c}
\hline\hline
[Fe/H] & $Z$ & $X$\\
\hline \hline
-2.00 & 0.00014 & 0.75115\\
-1.50 & 0.00043 & 0.75041\\
-1.35 &0.00061& 0.74996\\
-1.00 &0.00135& 0.74806\\
-0.50&0.00424&0.74073\\
-0.20 &0.00834& 0.73032\\
0.00 &0.01300& 0.71847\\
\hline
\end{tabular}
\begin{tablenotes}
	\small
	\item $^{*}$ The $Z$ and $X$ values are estimated from the $\mathrm{[Fe/H]}$ values by assuming the primordial helium value of 0.2485 from the WMAP CMB observations \citep{hinshaw2013} and the helium enrichment parameter value of 1.54 \citep{asplund2009}. The solar mixture is adopted from \citet{asplund2009}.
\end{tablenotes}
\label{tab:composition}
\end{table}

\item Stellar mass ($M$)
\begin{enumerate}
\item Low-mass range = 0.5M$_{\odot}$, 0.55M$_{\odot}$, 0.6M$_{\odot}$
\item High-mass range = 0.65M$_{\odot}$, 0.7M$_{\odot}$, 0.75M$_{\odot}$, 0.8M$_{\odot}$
\end{enumerate}

\item Stellar luminosity ($L$)
\begin{enumerate}
\item For low-mass range = 50L$_{\odot}$ to 200L$_{\odot}$, in steps of 50L$_{\odot}$
\item For high-mass range = 50L$_{\odot}$ to 300L$_{\odot}$, in steps of 50L$_{\odot}$
\end{enumerate}

\item Effective temperature ($T_{\rm{eff}}$) =  4000K to 8000K, in steps of 50K.
\end{enumerate}

\begin{figure}
\centering
\includegraphics[scale = 1]{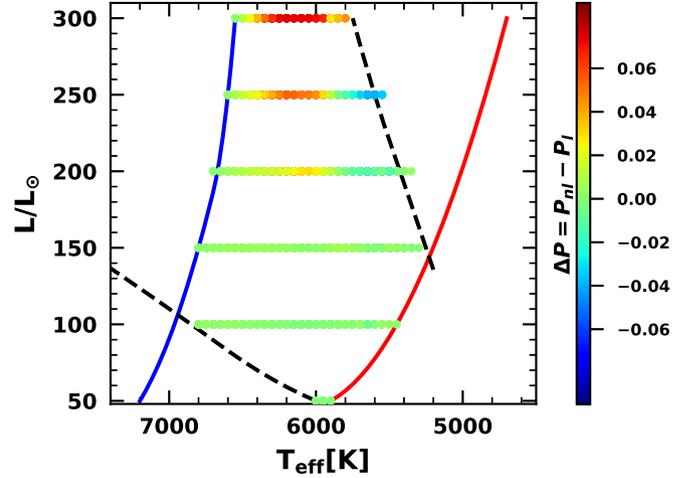}
\caption{The Hertzsprung-Russell diagram of the grid of computed BL~Her models with the convection parameter set A. The edges of the instability strip are estimated from the linear stability analysis and are shown with the blue and red solid lines. The dashed lines are lines of constant fundamental mode period (linear value) equal to 1 day (bottom) and 4 days (top). The dots represent the BL~Her models used in this analysis while the colourbar shows the difference between the non-linear and linear period of the computed models, i.e., $\Delta P = P_{nl}-P_l$. The mean difference between the linear and non-linear periods in our models is $\sim$0.02 days.}
\label{fig:HRD}
\end{figure}

This results in a combination of 20412 models per convective parameter set. BL~Her stars belong to the low-mass population with masses $\sim 0.5 M_{\odot}-0.6 M_{\odot}$ \citep{bhardwaj2020a}. However, we also explore the possibility of higher mass BL~Her stars in this study. In his survey of non-linear convective T2C models, \citet{smolec2016} computed a grid using $0.6 M_{\odot}$ and $0.8 M_{\odot}$. The effective temperature range chosen in the present study is much broader than the actual width of the instability strip to accurately estimate the edges of the instability strip. \textsc{mesa-rsp} may be used to compute models where the structure of the stellar envelope determines the pulsations \citep{paxton2019}, without taking into consideration the detailed structure of the core. We begin with a computation of linear properties of the models with the same method as described in \citet{smolec2016} and \citet{paxton2019}. To this end, equilibrium static models are constructed with $ZXMLT_{\rm{eff}}$ as the input stellar parameters and their linear stability analysis is conducted, which yields linear periods of the radial pulsation modes and their growth rates. The latter may be used to delineate the boundaries of the instability strip. The static models also serve as input for non-linear model integration. Fig.~\ref{fig:HRD} shows the HRD of the grid of computed BL~Her models with the convection parameter set A, showing the edges of the instability strip and the lines of constant fundamental mode period (linear value) equal to 1 and 4 days. The other convective parameter sets (B, C and D) show similar HRDs. A future paper will investigate linear results in greater detail.

\begin{figure*}
\centering
\includegraphics[scale = 0.5]{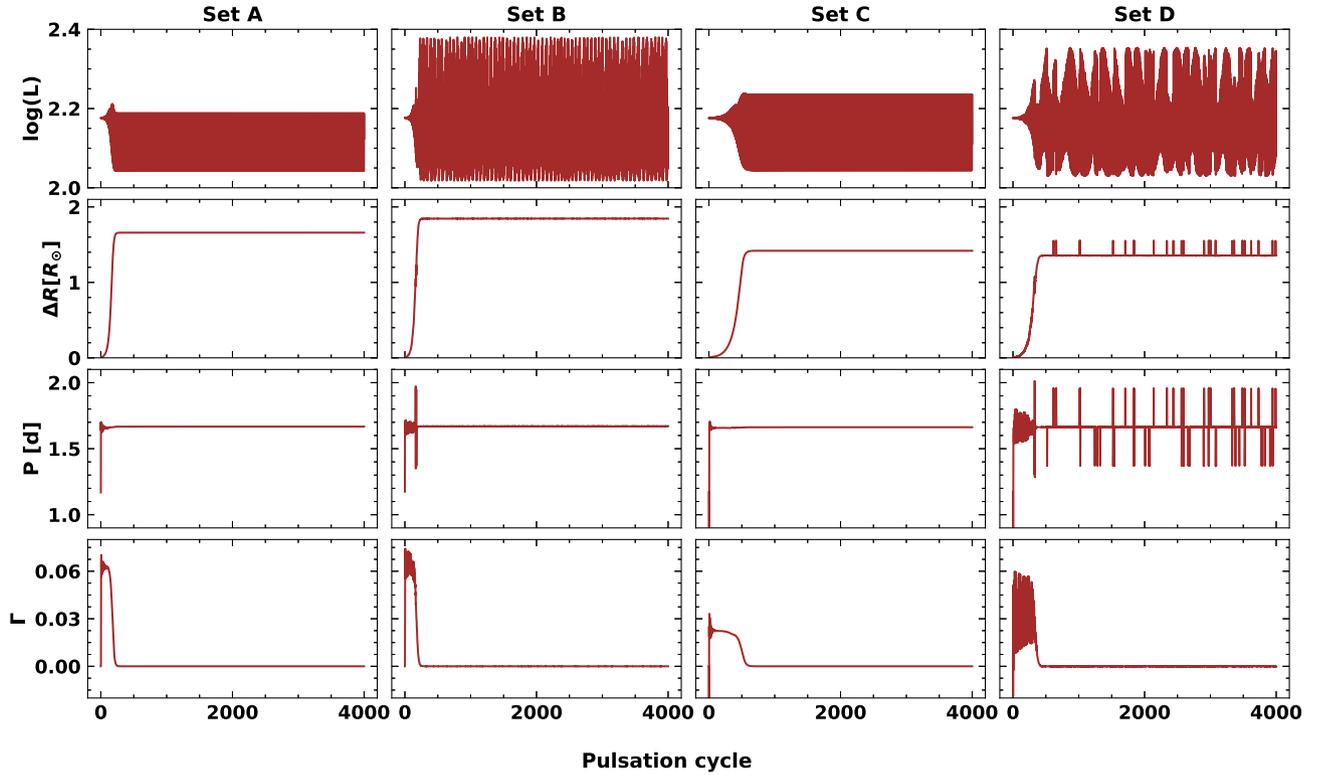}
\caption{Luminosity $\log(L)$, amplitude of radius variation $\Delta R$, period $P$ and fractional growth rate $\Gamma$ during 4000-cycle integrations of the BL~Her model with input parameters, $Z=0.013$, $X=0.71847$, $M=0.7$M$_{\odot}$, $L=150$L$_{\odot}$, $T=6350$K. The models with the same input parameters computed using sets A, B and C satisfy the condition of full-amplitude stable pulsation ($\Delta R$, $P$ and $\Gamma$ do not vary by more than 0.01 over the last $\sim$100-cycles of the total 4000-cycle integrations) and are accepted for the analysis while the model computed using Set D is rejected.}
\label{fig:stablepulsations}
\end{figure*}

\subsection{Non-Linear computations}

We proceed with the non-linear computations for models that have positive growth rates of the radial fundamental mode and linear periods between 0.8 and 4.2 days. The period range is larger than considered for BL~Her stars due to non-linear period changes. While non-linear period shifts expected for BL~Her models are well below 0.05 days \citep[see Fig. 6 in][]{smolec2016}, the changes of up to 0.2 days are noted for more luminous type~II Cepheid models. Finally, we only select models with non-linear periods between 1 and 4 days as BL~Her models, considering the typical period range of BL~Her stars \citep{soszynski2018}. We also confirm that non-linear period shifts are below 0.09 days in our models, with a mean non-linear period shift of $\sim$0.02 days.

Note that the quantities that enter into energy and momentum equations of the \textsc{mesa-rsp} convection model depend on the free parameters described in Table~\ref{tab:convection}; \citet{paxton2019} found that the pulsation periods of the models weakly depend on the values of these free parameters. Therefore, a model with the same input parameters $ZXMLT_{\rm{eff}}$ may have different non-linear periods across the different sets of convection parameters.

The non-linear model integration is carried for 4000 pulsation cycles; the control used for this terminating condition is \texttt{RSP\_max\_num\_periods} in \textsc{mesa-rsp}. It is essential to check for full-amplitude stable pulsations of the models before obtaining the theoretical $PL$ and $PR$ relations. The model reaches full amplitude pulsation state when its kinetic energy per pulsation period remains constant. We can quantify that using the fractional growth of the kinetic energy per pulsation period $\Gamma$ which should approach zero once stable pulsation state is reached. These pulsations may show some irregularities, e.g. period doubling effect or chaotic oscillations \citep{smolec2012a, smolec2012b, smolec2014}. Since we are interested in stable single periodic oscillation, we also investigate whether the amplitude of radius variation $\Delta R$ and the pulsation period $P$ computed on a cycle to cycle basis is stable. For our study, the condition of full-amplitude stable pulsation is satisfied and the model is accepted when its $\Delta R$, $P$ and $\Gamma$ do not vary by more than 0.01 over the last $\sim$100-cycles of the total 4000-cycle integrations, failing which the model is rejected from further analysis. Fig.~\ref{fig:stablepulsations} shows an example for one particular model- while sets A, B and C exhibit full-amplitude stable pulsations and are accepted, the set D model is most likely chaotic and is rejected from our analysis. The number of models accepted in sets B and D is much less than in sets A and C. The sets B and D include radiative cooling with the parameter $\gamma_r = 2\sqrt{3}$, while this parameter is set to zero in the sets A and C. Table~\ref{tab:number} summarises the number of BL~Her models for each convection parameter set.

We note that the higher stellar mass (>0.6$M_{\odot}$) and lower metallicity (Z=0.00014) considered in our work is typical of Zero Age Horizontal Branch or evolved RR~Lyrae stars. However, as discussed in \citet{braga2020}, the separation between RR~Lyrae and T2Cs is a long-standing problem. A threshold of period $\sim$1 day typically separates the two classes of pulsating variable stars \citep{soszynski2008,soszynski2014}. \citet{braga2020} report the star V92 in $\omega$-Cen (with a period of 1.3 days) to be a candidate RR~Lyrae star because its core is likely in the helium-burning phase. Given the evidence, it is rather difficult to separate completely unambiguously the two different classes of pulsating variables based on their chemical composition. It might be possible to separate RR~Lyrae and BL~Her stars based on their evolutionary status, although this is not easily done for field stars. However, in this work, we use the conventional classification for BL~Her variables as population~II stars with pulsation period between 1 and 4 days \citep{soszynski2018}.

\begin{table}
\caption{A summary of the number of BL~Her models in each convection parameter set finally used in the analysis.}
\centering
\scalebox{0.85}{
\begin{tabular}{l c c c c}
\hline\hline
Condition & Set A & Set B & Set C & Set D\\
\hline \hline
Total $ZXMLT_{\rm{eff}}$ combinations & 20412 & 20412 & 20412 & 20412\\
(Models computed in the linear grid) & & & & \\
\hline
Models with positive growth rate of the F-mode & 4481 & 4356 & 4061 & 4192\\
and with linear period: $0.8 \leq P \leq 4.2$ & & & & \\
(Models computed in the non-linear grid) & & & & \\
\hline
Models with non-linear period: $1 \leq P \leq 4$ & 4049 & 3854 & 3629 & 3678\\
\hline
Models with full-amplitude stable pulsation$^\dagger$ &3266  &2260  &2632  &2122\\
\hline
\end{tabular}}
\begin{tablenotes}
	\small
	\item $^{\dagger}$ Satisfies the condition that the amplitude of radius variation $\Delta R$, period $P$ and fractional growth rate $\Gamma$ do not vary by more than 0.01 over the last $\sim$100-cycles of the total 4000-cycle integrations. For a clear, pictorial representation of full-amplitude stable pulsation, the reader may refer to Fig.~\ref{fig:stablepulsations}.
	
\end{tablenotes}
\label{tab:number}
\end{table}

\subsection{Processing the data}

The details on the transformation of bolometric light curves into optical and NIR bands is given in \citet{paxton2018} and is briefly summarised here. The luminosity $\rm{\log_{10} (L/L_{\odot})}$ obtained from the non-linear computations of the models is converted to the absolute bolometric magnitude ($\rm{M_{bol}}$) of the model using:
\begin{equation}
\rm{M_{bol} = M_{bol,\odot} - 2.5\log_{10} (L/L_{\odot})},
\end{equation}
where $\rm{M_{bol,\odot}} = 4.74$ \citep{mamajek2015} is the absolute bolometric magnitude of the Sun. The absolute bolometric magnitude is then transformed into the absolute magnitude $\rm{M_\lambda}$ in a given band $\lambda$ using:
\begin{equation}
\rm{M_\lambda=M_{bol}-BC_\lambda},
\end{equation}
where $\rm{BC_\lambda}$ is the bolometric correction for band $\lambda$. \textsc{mesa} provides pre-computed bolometric correction tables where the bolometric correction is defined as a function of the stellar photosphere. For given stellar photosphere parameters, the bolometric correction table is interpolated over $\log(T)$, $\log(g)$ and the metallicity [M/H] within the parameter range of that table. We use the pre-processed table from \citet{lejeune1998} which provides bolometric corrections over the parameter range $2000 \leq T_{\rm{eff}} \rm{(K)} \leq 50,000$, $-1.02 \leq \log (g) \rm{(cm s^{-2})} \leq 5.0$, and $-5.0 \leq [M/H] \leq 1.0$ and for the Johnson-Cousins-Class bands $UBVR_cI_cJHKLL'M$. The minimal impact of adopted transformations on the mean-light properties at wavelengths longer than $V$-band is discussed in Appendix~\ref{sec:appendix}.

\begin{figure}
\centering
\includegraphics[scale = 1]{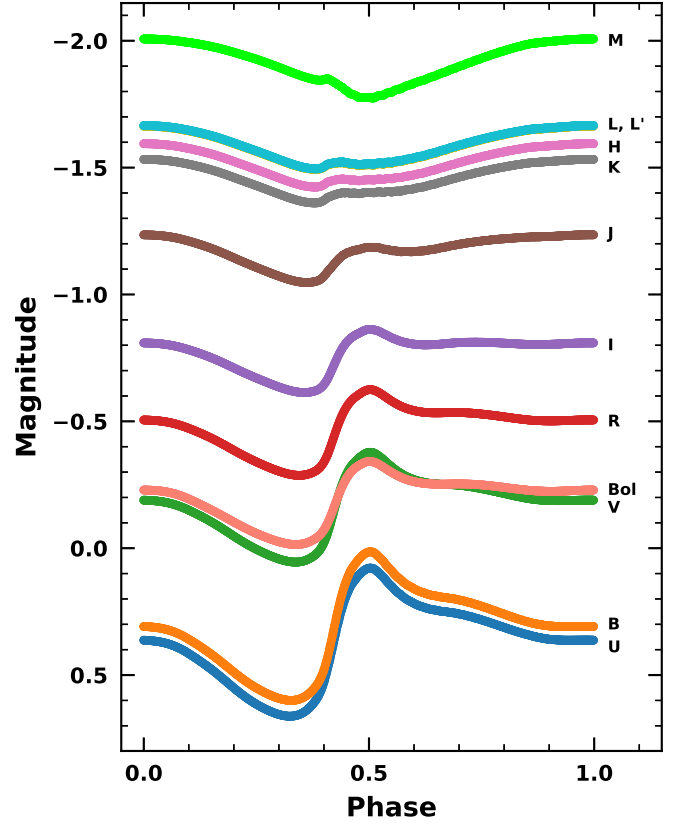}
\caption{An example of the multi-wavelength light-curves for the BL~Her model with input parameters, $Z=0.00424$, $X=0.74073$, $M=0.80$M$_{\odot}$, $L=100$L$_{\odot}$, $T=6050$K using convective parameter set A.}				
\label{fig:LC}
\end{figure}

The multi-wavelength theoretical light-curves of the accepted models are fitted with the Fourier sine series \citep[see example,][]{deb2009,bhardwaj2015,das2018} of the form:
\begin{equation}
m(x) = m_0 + \sum_{k=1}^{N}A_k \sin(2 \pi kx+\phi_k),
\label{eq:fourier}
\end{equation}

\noindent where $x$ is the pulsation phase, $m_0$ is the mean magnitude and $N$ is the order of the fit ($N = 20$). Table~\ref{tab:allmodels} summarises the input stellar parameters of the BL~Her models used in this analysis, alongwith the multi-wavelength absolute mean magnitudes obtained from the Fourier fitting. An example of light-curves for a BL~Her model obtained using \textsc{mesa-rsp} over multiple wavelengths is presented in Fig.~\ref{fig:LC}.

\begin{table*}
\caption{A summary of the BL~Her models used in this analysis computed using \textsc{mesa-rsp}. The columns provide the input parameters ($ZXMLT_{\rm{eff}}$), the convection parameter set used, logarithmic pulsation period, logarithmic radius and the absolute mean magnitudes in the bands $UBVR_cI_cJHKLL'M$ and the bolometric.}
\centering
\scalebox{0.8}{
\begin{tabular}{c c c c c c c c c c c c c c c c c c c c}
\hline\hline
$Z$ & $X$ & $M/M_{\odot}$ & $L/L_{\odot}$ & $\rm{T_{eff} (K)}$ & \makecell{Convection\\ Set} & $\log(P)$ & $\log(R)$ & $M_U$ & $M_B$ & $M_V$ & $M_R$ & $M_I$ & $M_J$ & $M_H$ & $M_K$ & $M_L$ & $M_{L'}$ & $M_M$ & $M_{Bol}$\\
\hline \hline
0.00014&	0.75115&	0.50&	50&	5950&	A&	0.011&	0.826&	0.971&	1.087&	0.660&	0.350&	0.033&	-0.412&	-0.793&	-0.732&	-0.867&	-0.870&	-1.242&	0.493\\
0.00014&	0.75115&	0.50&	100&	5650&	A&	0.367&	1.024&	0.382&	0.452&	-0.051&	-0.411&	-0.774&	-1.259&	-1.700&	-1.627&	-1.782&	-1.787&	-2.223&	-0.257\\
...& ...& ...& ...& ...& ...& ...& ...& ...& ...& ...& ...& ...& ...& ...& ...& ...& ...& ...& ...\\
0.00014&	0.75115&	0.50&	100&	5700&	B&	0.355&	1.018&	0.351&	0.428&	-0.060&	-0.413&	-0.770&	-1.246&	-1.677&	-1.606&	-1.757&	-1.762&	-2.187&	-0.258\\
0.00014&	0.75115&	0.50&	100&	5750&	B&	0.336&	1.013&	0.342&	0.417&	-0.064&	-0.411&	-0.761&	-1.230&	-1.653&	-1.583&	-1.733&	-1.737&	-2.155&	-0.256\\
...& ...& ...& ...& ...& ...& ...& ...& ...& ...& ...& ...& ...& ...& ...& ...& ...& ...& ...& ...\\
0.00014&	0.75115&	0.50&	50&	5850&	C&	0.059&	0.857&	1.036&	1.159&	0.689&	0.346&	-0.003&	-0.463&	-0.879&	-0.813&	-0.957&	-0.961&	-1.370&	0.493\\
0.00014&	0.75115&	0.50&	50&	5900&	C&	0.038&	0.848&	1.015&	1.137&	0.681&	0.348&	0.009&	-0.446&	-0.850&	-0.786&	-0.927&	-0.931&	-1.328&	0.493\\
...& ...& ...& ...& ...& ...& ...& ...& ...& ...& ...& ...& ...& ...& ...& ...& ...& ...& ...& ...\\
0.00014&	0.75115&	0.50&	50&	5900&	D&	0.039&	0.848&	1.012&	1.135&	0.680&	0.347&	0.008&	-0.445&	-0.849&	-0.786&	-0.926&	-0.930&	-1.325&	0.493\\
0.00014&	0.75115&	0.50&	50&	5950&	D&	0.019&	0.839&	0.994&	1.114&	0.672&	0.349&	0.020&	-0.430&	-0.823&	-0.761&	-0.899&	-0.902&	-1.287&	0.493\\
...& ...& ...& ...& ...& ...& ...& ...& ...& ...& ...& ...& ...& ...& ...& ...& ...& ...& ...& ...\\
\hline
\end{tabular}
}
\begin{tablenotes}
	\small
	\item \textit{Note:} This table is available entirely in a machine-readable form in the online journal as supporting information.
\end{tablenotes}
\label{tab:allmodels}
\end{table*}

\section{Period-luminosity relations}
\label{sec:PL}

The mean magnitudes obtained from Fourier fitting of the theoretical light-curves of the BL~Her models are used to derive multi-wavelength $PL$ relations of the mathematical form:
\begin{equation}
M_\lambda=a\log(P)+b,
\end{equation}
where, $M_\lambda$ refers to the absolute magnitude in a given band, $\lambda$.

\subsection{Effect of convection parameters on $PL$ relations}

To study the effect of convection parameters on the $PL$ relations, we use the standard $t$-test to check the statistical equivalence of the slopes from $PL$ relations of the BL~Her models obtained using different convective parameter sets. A detailed description of the test is provided in \citet{ngeow2015} and \citet{das2020} and is summarised here briefly. The $T$ statistic to compare slopes, $\hat{W}$ of two linear regressions with sample sizes, $n$ and $m$, respectively is defined as:
\begin{equation}
T=\frac{\hat{W}_n-\hat{W}_m}{\sqrt{\mathrm{Var}(\hat{W}_n)+\mathrm{Var}(\hat{W}_m)}},
\label{eq:ttest}
\end{equation}
where $\mathrm{Var}(\hat{W})$ is the variance of the slope. We reject the null hypothesis of equivalent slopes if $T>t_{\alpha/2,\nu}$ or the probability of the observed value of the $T$ statistic is $p<0.05$. $t_{\alpha/2,\nu}$ is the critical value under the two-tailed $t$-distribution with 95\% confidence limit ($\alpha$=0.05) and degrees of freedom, $\nu=n+m-4$.

Table~\ref{tab:PL} lists the statistical comparison of the multi-wavelength $PL$ relation slopes of the BL~Her models with respect to literature values. The BL~Her models with radiative cooling (sets B and D) exhibit statistically similar $PL$ slopes at any given wavelength, except in the $U$-band while those without radiative cooling (sets A and C) have statistically similar $PL$ slopes only in the $HLL'$ bands. BL~Her models computed with different convection parameters do show differences in the slopes of the $PL$ relations at a given wavelength. However, most empirical $PL$ relations are consistent with theoretical $PL$ relations based on any set of convection parameter. The $PL$ slopes obtained for the BL~Her stars in the Galactic bulge by \citet{bhardwaj2017c} and in the SMC by \citet{matsunaga2009} are statistically similar with those obtained from the BL~Her models using all four sets of convection parameters and across $JHK_S$ bands. Our models in all four convection parameter sets show statistically different $PL$ slopes from the $PL$ slope obtained by \citet{matsunaga2006} for BL~Her stars in the Globular clusters in the $J$-band; however, they exhibit statistically similar $PL$ slopes in the $HK_S$ bands. For the BL~Her stars in the LMC, $PL$ relations in the $J$-band show similar slopes between empirical data and models using all four sets of convection parameters; however, $PL$ slopes of the models with sets B and D seem to match better with empirical relations in the $HK_S$ bands. The bolometric $PL$ relations obtained by \citet{groenewegen2017b} for BL~Her stars in the Magellanic Clouds show statistically similar slopes with models of different convection parameter sets, except for models computed with set A when the observed BL~Her stars in the LMC and the SMC are considered together. The fact that the $PL$ slopes are statistically different between different sets of models, but empirically determined $PL$ slopes are still compatible with all four sets of model $PL$ slopes tells us that the models have lower uncertainty and scatter. Larger and more precise observational datasets are necessary to constrain the best fitting model parameters. From Table~\ref{tab:PL}, we find that the $PL$ slopes from the BL~Her models become steeper with increasing wavelengths. This is similar to RR Lyrae as shown empirically in \citet{beaton2018b, neeley2017, bhardwaj2020b}. The dispersion in the theoretical $PL$ relations for BL~Hers decreases significantly on moving from optical to infrared wavelengths and becomes statistically similar for wavelengths longer than $H$-band. A similar decrease of intrinsic dispersion of the $PL$ relations when changing from the optical to infrared bands has been reported by \citet{neeley2017} based on RR~Lyrae models from \citet{marconi2015}, and is also seen in empirical $PL$ relations of RR~Lyrae \citep[e.g.][]{bhardwaj2020b}. This trend in dispersion is expected because of the stronger temperature sensitivity of the bolometric correction in the near-infrared, resulting in brighter magnitudes of BL~Hers at cooler effective temperatures \citep{bono2003} and a marginal effect of the intrinsic temperature width of the instability strip on the infrared $PL$ relations. In addition, the width of the instability strip itself decreases at longer wavelengths, resulting in smaller dispersion of $PL$ relations \citep{catelan2004, madore2012, marconi2015}. We also observe that the models computed using set B have the smallest dispersion in their $PL$ relations in all the bands.

\begin{table*}
\caption{Comparison of the slopes of the period-luminosity relations for BL~Her stars of the mathematical form $M_\lambda=a\log(P)+b$. $N$ is the total number of stars and the number in parentheses indicates the number of outliers. |$T$| represents the observed value of the $t$-statistic, and $p(t)$ gives the probability of acceptance of the null hypothesis (equal slopes). The bold-faced entries indicate that the null hypothesis of the equivalent $PL$ slopes can be rejected.}
\centering
\scalebox{0.8}{
\begin{tabular}{c c c c c c c c c c c c}
\hline\hline
Band & Source & $a$ & $b$ & $\sigma$ & $N$ & Reference${^\ddagger}$ & \makecell[c]{Theoretical/ \\ Empirical} & \multicolumn{4}{c}{(|$T$|, $p(t)$) w.r.t.}\\
& & & & & & & & Set A & Set B & Set C & Set D\\
\hline \hline
U & $\rm{Z_{all}}$ (Set A) &-0.841$\pm$0.044&0.185$\pm$0.015&0.391&3266& TW & Theoretical & ... & ... & ... & ...\\
U & $\rm{Z_{all}}$ (Set B) &-0.596$\pm$0.046&0.215$\pm$0.015&0.353&2260& TW & Theoretical & \textbf{(3.846,0.0)}& ... & ... & ...\\
U & $\rm{Z_{all}}$ (Set C) &-0.422$\pm$0.051&0.298$\pm$0.018&0.428&2632& TW & Theoretical & \textbf{(6.219,0.0)} & \textbf{(2.512,0.006)} & ... & ...\\
U & $\rm{Z_{all}}$ (Set D) &-0.369$\pm$0.053&0.309$\pm$0.018&0.409&2122& TW & Theoretical & \textbf{(6.851,0.0)} & \textbf{(3.213,0.001)} & (0.722,0.235) & ...\\
\hline
B & $\rm{Z_{all}}$ (Set A) &-1.166$\pm$0.04&0.187$\pm$0.014&0.351&3266& TW & Theoretical & ... & ... & ... & ...\\
B & $\rm{Z_{all}}$ (Set B) &-0.896$\pm$0.041&0.209$\pm$0.013&0.311&2260& TW & Theoretical & \textbf{(4.764,0.0)}& ... & ... & ...\\
B & $\rm{Z_{all}}$ (Set C) &-0.942$\pm$0.043&0.33$\pm$0.015&0.359&2632& TW & Theoretical & \textbf{(3.843,0.0)} & (0.791,0.214) & ... & ...\\
B & $\rm{Z_{all}}$ (Set D) &-0.805$\pm$0.044&0.324$\pm$0.015&0.339&2122& TW & Theoretical & \textbf{(6.1,0.0)} & (1.508,0.066) & \textbf{(2.235,0.013)} & ...\\
\hline
V & $\rm{Z_{all}}$ (Set A) &-1.616$\pm$0.032&-0.14$\pm$0.011&0.284&3266& TW & Theoretical & ... & ... & ... & ...\\
V & $\rm{Z_{all}}$ (Set B) &-1.374$\pm$0.034&-0.134$\pm$0.011&0.256&2260& TW & Theoretical & \textbf{(5.221,0.0)}& ... & ... & ...\\
V & $\rm{Z_{all}}$ (Set C) &-1.487$\pm$0.034&-0.031$\pm$0.012&0.288&2632& TW & Theoretical & \textbf{(2.761,0.003)}& \textbf{(2.343,0.01)} & ... & ...\\
V & $\rm{Z_{all}}$ (Set D) &-1.337$\pm$0.036&-0.043$\pm$0.012&0.275&2122& TW & Theoretical & \textbf{(5.829,0.0)} & (0.757,0.225) & \textbf{(3.023,0.001)} & ...\\
\hline
R & $\rm{Z_{all}}$ (Set A) &-1.853$\pm$0.028&-0.364$\pm$0.01&0.248&3266& TW & Theoretical & ... & ... & ... & ...\\
R & $\rm{Z_{all}}$ (Set B) &-1.634$\pm$0.03&-0.365$\pm$0.01&0.227&2260& TW & Theoretical & \textbf{(5.357,0.0)}& ... & ... & ...\\
R & $\rm{Z_{all}}$ (Set C) &-1.739$\pm$0.03&-0.281$\pm$0.01&0.254&2632& TW & Theoretical & \textbf{(2.774,0.003)} & \textbf{(2.463,0.007)} & ... & ...\\
R & $\rm{Z_{all}}$ (Set D) &-1.6$\pm$0.032&-0.293$\pm$0.011&0.243&2122& TW & Theoretical & \textbf{(5.995,0.0)} & (0.783,0.217) & \textbf{(3.169,0.001)} & ...\\
\hline
I & $\rm{Z_{all}}$ (Set A) &-2.043$\pm$0.025&-0.592$\pm$0.008&0.219&3266& TW & Theoretical & ... & ... & ... & ...\\
I & $\rm{Z_{all}}$ (Set B) &-1.848$\pm$0.027&-0.599$\pm$0.009&0.203&2260& TW & Theoretical & \textbf{(5.386,0.0)}& ... & ... & ...\\
I & $\rm{Z_{all}}$ (Set C) &-1.932$\pm$0.027&-0.534$\pm$0.009&0.226&2632& TW & Theoretical & \textbf{(3.05,0.001)} & \textbf{(2.223,0.013)} & ... & ...\\
I & $\rm{Z_{all}}$ (Set D) &-1.81$\pm$0.028&-0.545$\pm$0.01&0.218&2122& TW & Theoretical & \textbf{(6.225,0.0)} & (0.976,0.165) & \textbf{(3.129,0.001)} & ...\\
\hline
J & $\rm{Z_{all}}$ (Set A) &-2.303$\pm$0.021&-0.914$\pm$0.007&0.186&3266& TW & Theoretical & ... & ... & ... & ...\\
J & $\rm{Z_{all}}$ (Set B) &-2.131$\pm$0.023&-0.928$\pm$0.008&0.177&2260& TW & Theoretical & \textbf{(5.505,0.0)}& ... & ... & ...\\
J & $\rm{Z_{all}}$ (Set C) &-2.239$\pm$0.023&-0.877$\pm$0.008&0.19&2632& TW & Theoretical & \textbf{(2.067,0.019)} & \textbf{(3.332,0.0)} & ... & ...\\
J & $\rm{Z_{all}}$ (Set D) &-2.122$\pm$0.024&-0.89$\pm$0.008&0.187&2122& TW & Theoretical & \textbf{(5.617,0.0)} & (0.243,0.404) & \textbf{(3.497,0.0)} & ...\\

J & Globular clusters & -2.959$\pm$0.313 & -1.541$\pm$0.041 (@0.3)$^{*}$ & 0.11 & 7 & M06 & Empirical & \textbf{(2.091,0.018)} & \textbf{(2.638,0.004)} & \textbf{(2.294,0.011)} & \textbf{(2.666,0.004)}\\
J & Galactic bulge & -2.387$\pm$0.164 & 11.393$\pm$0.132 (@1.0)$^{\dagger}$ & 0.347 & 106 & B17b & Empirical & (0.508,0.306) & (1.546,0.061) & (0.894,0.186) & (1.599,0.055)\\
J & LMC & -2.164$\pm$0.240 & 17.131$\pm$0.038 (@0.3)$^{*}$ & 0.25 & 55 & M09 & Empirical & (0.577,0.282) & (0.137,0.446) & (0.311,0.378) & (0.174,0.431)\\
J & LMC & -2.294$\pm$0.153 & 15.375$\pm$0.113 (@1.0)$^{\dagger}$ & 0.202 & 55 & B17a & Empirical & (0.058,0.477) & (1.054,0.146) & (0.355,0.361) & (1.111,0.133)\\
J & SMC (IRSF only) & -2.545$\pm$0.764 & 17.393$\pm$0.112 (@0.3)$^{*}$ & 0.41 & 15 & M11 & Empirical & (0.317,0.376) & (0.542,0.294) & (0.4,0.345) & (0.553,0.29)\\
J & SMC (IRSF+NTT) & -2.690$\pm$0.488 & 17.325$\pm$0.069 (@0.3)$^{*}$ & 0.36 & 31 & M11 & Empirical & (0.792,0.214) & (1.144,0.126) & (0.923,0.178) & (1.163,0.122)\\

\hline
H & $\rm{Z_{all}}$ (Set A) &-2.57$\pm$0.018&-1.17$\pm$0.006&0.157&3266& TW & Theoretical & ... & ... & ... & ...\\
H & $\rm{Z_{all}}$ (Set B) &-2.429$\pm$0.02&-1.192$\pm$0.007&0.154&2260& TW & Theoretical & \textbf{(5.236,0.0)}& ... & ... & ...\\
H & $\rm{Z_{all}}$ (Set C) &-2.529$\pm$0.019&-1.162$\pm$0.007&0.16&2632& TW & Theoretical & (1.568,0.058) & \textbf{(3.587,0.0)} & ... & ...\\
H & $\rm{Z_{all}}$ (Set D) &-2.432$\pm$0.021&-1.175$\pm$0.007&0.162&2122& TW & Theoretical & \textbf{(5.048,0.0)} & (0.081,0.468) & \textbf{(3.439,0.0)} & ...\\

H & Globular clusters & -2.335$\pm$0.335 & -1.847$\pm$0.044 (@0.3)$^{*}$ & 0.12 & 7 & M06 & Empirical & (0.7,0.242) & (0.28,0.39) & (0.578,0.282) & (0.289,0.386)\\
H & Galactic bulge & -2.591$\pm$0.163 & 11.019$\pm$0.130 (@1.0)$^{\dagger}$ & 0.353 & 104 & B17b & Empirical& (0.128,0.449) & (0.986,0.162) & (0.378,0.353) & (0.967,0.167)\\
H & LMC & -2.259$\pm$0.248 & 16.857$\pm$0.039 (@0.3)$^{*}$ & 0.26 & 54 & M09 & Empirical & (1.251,0.106) & (0.683,0.247) & (1.086,0.139) & (0.695,0.244)\\
H & LMC & -2.088$\pm$0.214 & 15.218$\pm$0.163 (@1.0)$^{\dagger}$ & 0.296 & 52 & B17a & Empirical & \textbf{(2.244,0.012)} & (1.587,0.056) & \textbf{(2.053,0.02)} & (1.6,0.055)\\
H & SMC (IRSF only) & -2.765$\pm$0.731 & 17.080$\pm$0.108 (@0.3)$^{*}$ & 0.40 & 15 & M11 & Empirical& (0.267,0.395) & (0.459,0.323) & (0.323,0.373) & (0.455,0.325)\\
\hline
K & $\rm{Z_{all}}$ (Set A) &-2.528$\pm$0.018&-1.124$\pm$0.006&0.16&3266& TW & Theoretical & ... & ... & ... & ...\\
K & $\rm{Z_{all}}$ (Set B) &-2.383$\pm$0.021&-1.144$\pm$0.007&0.157&2260& TW & Theoretical & \textbf{(5.308,0.0)}& ... & ... & ...\\
K & $\rm{Z_{all}}$ (Set C) &-2.483$\pm$0.02&-1.112$\pm$0.007&0.164&2632& TW & Theoretical & \textbf{(1.7,0.045)} & \textbf{(3.526,0.0)} & ... & ...\\
K & $\rm{Z_{all}}$ (Set D) &-2.383$\pm$0.021&-1.125$\pm$0.007&0.165&2122& TW & Theoretical & \textbf{(5.194,0.0)} & (0.008,0.497) & \textbf{(3.453,0.0)} & ...\\

K$_S$ & Globular clusters & -2.294$\pm$0.294 & -1.864$\pm$0.039 (@0.3)$^{*}$ & 0.10 & 7 & M06 & Empirical & (0.794,0.214) & (0.302,0.381) & (0.641,0.261) & (0.302,0.381)\\
K$_S$ & Galactic bulge & -2.362$\pm$0.170 & 11.071$\pm$0.133 (@1.0)$^{\dagger}$ & 0.294 & 108 & B17b & Empirical& (0.971,0.166) & (0.123,0.451) & (0.707,0.24) & (0.123,0.451)\\
K$_S$ & LMC & -1.992$\pm$0.278 & 16.733$\pm$0.040 (@0.3)$^{*}$ & 0.26 & 47 & M09 & Empirical& \textbf{(1.924,0.027)} & (1.402,0.081) & \textbf{(1.762,0.039)} & (1.402,0.081)\\
K$_S$ & LMC & -2.083$\pm$0.154 & 15.162$\pm$0.114 (@1.0)$^{\dagger}$ & 0.262 & 47 & B17a & Empirical & \textbf{(2.87,0.002)} & \textbf{(1.93,0.027)} & \textbf{(2.576,0.005)} & \textbf{(1.93,0.027)}\\
K$_S$ & SMC (IRSF only)& -2.096$\pm$0.732 & 16.933$\pm$0.104 (@0.3)$^{*}$ & 0.37 & 13 & M11 & Empirical& (0.59,0.278) & (0.392,0.348) & (0.528,0.299) & (0.392,0.348)\\
K$_S$ & SMC (IRSF+NTT) & -2.553$\pm$0.444 & 16.924$\pm$0.061 (@0.3)$^{*}$ & 0.32 & 29 & M11 & Empirical& (0.056,0.478) & (0.382,0.351) & (0.157,0.438) & (0.382,0.351)\\

\hline
L & $\rm{Z_{all}}$ (Set A) &-2.599$\pm$0.017&-1.231$\pm$0.006&0.155&3266& TW & Theoretical & ... & ... & ... & ...\\
L & $\rm{Z_{all}}$ (Set B) &-2.461$\pm$0.02&-1.253$\pm$0.007&0.153&2260& TW & Theoretical & \textbf{(5.195,0.0)}& ... & ... & ...\\
L & $\rm{Z_{all}}$ (Set C) &-2.56$\pm$0.019&-1.226$\pm$0.007&0.158&2632& TW & Theoretical & (1.509,0.066) & \textbf{(3.607,0.0)} & ... & ...\\
L & $\rm{Z_{all}}$ (Set D) &-2.464$\pm$0.021&-1.238$\pm$0.007&0.16&2122& TW & Theoretical & \textbf{(4.977,0.0)} & (0.113,0.455) & \textbf{(3.427,0.0)} & ...\\
\hline
L' & $\rm{Z_{all}}$ (Set A) &-2.599$\pm$0.017&-1.234$\pm$0.006&0.155&3266& TW & Theoretical & ... & ... & ... & ...\\
L' & $\rm{Z_{all}}$ (Set B) &-2.461$\pm$0.02&-1.256$\pm$0.007&0.153&2260& TW & Theoretical & \textbf{(5.173,0.0)}& ... & ... & ...\\
L' & $\rm{Z_{all}}$ (Set C) &-2.561$\pm$0.019&-1.229$\pm$0.007&0.158&2632& TW & Theoretical & (1.477,0.07) & \textbf{(3.617,0.0)} & ... & ...\\
L' & $\rm{Z_{all}}$ (Set D) &-2.465$\pm$0.021&-1.241$\pm$0.007&0.16&2122& TW & Theoretical & \textbf{(4.933,0.0)} & (0.134,0.447) & \textbf{(3.416,0.0)} & ...\\
\hline
M & $\rm{Z_{all}}$ (Set A) &-2.802$\pm$0.016&-1.492$\pm$0.006&0.144&3266& TW & Theoretical & ... & ... & ... & ...\\
M & $\rm{Z_{all}}$ (Set B) &-2.695$\pm$0.019&-1.519$\pm$0.006&0.145&2260& TW & Theoretical & \textbf{(4.277,0.0)}& ... & ... & ...\\
M & $\rm{Z_{all}}$ (Set C) &-2.759$\pm$0.018&-1.514$\pm$0.006&0.148&2632& TW & Theoretical & \textbf{(1.81,0.035)} & \textbf{(2.449,0.007)} & ... & ...\\
M & $\rm{Z_{all}}$ (Set D) &-2.692$\pm$0.02&-1.523$\pm$0.007&0.151&2122& TW & Theoretical & \textbf{(4.316,0.0)} & (0.118,0.453) & \textbf{(2.524,0.006)} & ...\\
\hline
Bolometric & $\rm{Z_{all}}$ (Set A) &-1.799$\pm$0.028&-0.181$\pm$0.01&0.253&3266& TW & Theoretical & ... & ... & ... & ...\\
Bolometric & $\rm{Z_{all}}$ (Set B) &-1.581$\pm$0.03&-0.18$\pm$0.01&0.231&2260& TW & Theoretical & \textbf{(5.245,0.0)}& ... & ... & ...\\
Bolometric & $\rm{Z_{all}}$ (Set C) &-1.693$\pm$0.031&-0.094$\pm$0.011&0.256&2632& TW & Theoretical & \textbf{(2.532,0.006)} & \textbf{(2.609,0.005)} & ... & ...\\
Bolometric & $\rm{Z_{all}}$ (Set D) &-1.559$\pm$0.032&-0.103$\pm$0.011&0.246&2122& TW & Theoretical & \textbf{(5.625,0.0)} & (0.512,0.304) & \textbf{(3.051,0.001)} & ...\\

Bolometric & LMC & -1.749$\pm$0.200 & 0.141$\pm$0.051 & 0.274 & 57(4) & G17 & Empirical & (0.248,0.402) & (0.831,0.203) & (0.277,0.391) & (0.938,0.174)\\
Bolometric & SMC & -0.691$\pm$0.717 & -0.250$\pm$0.176 & 0.302 & 15(2) & G17 & Empirical & (1.544,0.061) & (1.24,0.108) & (1.396,0.081) & (1.209,0.113)\\
Bolometric & MCs & -1.326$\pm$0.257 & -0.027$\pm$0.065 & 0.282 & 72(6) & G17 & Empirical & \textbf{(1.83,0.034)} & (0.986,0.162) & (1.418,0.078) & (0.9,0.184)\\
\hline
\end{tabular}}
\begin{tablenotes}
	\small
	\item ${^\ddagger}$ TW=This work; M06=\citet{matsunaga2006}; M09=\citet{matsunaga2009}; M11=\citet{matsunaga2011}; B17a=\citet{bhardwaj2017b}; B17b=\citet{bhardwaj2017c}; G17=\citet{groenewegen2017b}
	\item $^{*}$ Zero point at $\log(P)=0.3$    
	\item $^{\dagger}$ Zero point at $\log(P)=1.0$     
\end{tablenotes}
\label{tab:PL}
\end{table*}

\subsection{Effect of metallicity on $PL$ relations}
\label{sec:PLZ}

To quantify the effect of metallicity on the $PL$ relations, we obtain $PLZ$ relations for the BL~Her models of the mathematical form:
\begin{equation}
M_\lambda=a+b\log(P)+c\mathrm{[Fe/H]}
\end{equation}
The results of the $PLZ$ relations from the BL~Her models for different wavelengths using different convective parameter sets is summarised in Table~\ref{tab:PLZ}. The coefficients for the metallicity term from these relations suggest strong dependence of $PL$ relations on metallicity in $U$ and $B$ bands but only modest effect at longer wavelengths. This result holds true for the four different convective parameter sets A, B, C and D. The weak or no dependence of $PL$ relations on metallicity, especially at longer wavelengths is in agreement with earlier empirical evidence from \citet{matsunaga2006} and \citet{groenewegen2017b}. \citet{matsunaga2006} had studied the effect of metallicity on the $PL$ relations of T2Cs in the GGCs in the NIR $JHK_S$ bands while \citet{groenewegen2017b} had investigated the bolometric relations for T2Cs in the Magellanic Clouds. Table~\ref{tab:PLZ} also shows the theoretical $PLZ$ relations obtained by \citet{criscienzo2007} for BL~Her models in the $IJHK$ bands. However, we note here that \citet{criscienzo2007} use BL~Her models with $-2.62\; \mathrm{dex} \leq \mathrm{[Fe/H]} \leq -0.66\; \mathrm{dex}$ and $0.8 < \mathrm{P (days)} < 8$ in their study. 

The minimal dependence of metallicity on the $PL$ relation in the $V$-band onward is very interesting. A possible reason for significant metallicity dependence in $U$ and $B$-band could be that the effect of adopted model atmospheres on the transformations of bolometric light curves is significant at these wavelengths, as discussed in Appendix~\ref{sec:appendix}. To further investigate the dependence on metallicity, we separated models in low-metallicity ($Z=0.00014, 0.00043, 0.00061, 0.00135$) and high-metallicity ($Z=0.00424, 0.00834, 0.01300$) regime. The results of $PLZ$ relations for different convection sets are listed in Appendix~Tables~\ref{tab:B1} and \ref{tab:B2}. For the convection set A, we find that in the low-metallicity regime, only the $U$-band $PL$ relation displays a statistically significant dependence on metallicity. However, both $U$ and $B$-band $PL$ relations display a clear dependence on metallicity in the high-metallicity regime. The metallicity coefficient of $V$-band $PL$ relation is consistent with zero even for high metallicities. However, the $PL$ relations based on bolometric magnitudes show a marginal dependence on metallicity for convection set A which becomes consistent with zero for convection set D. This hints that the metallicity effects become significant in $U$ and $B$-bands because of the increasing sensitivity of bolometric corrections to metallicities at wavelengths shorter than $V$-band \citep{gray2005,kudritzki2008}. Furthermore, the zero-point of $PL$ relations based on $V$-band and bolometric magnitudes are similar but the difference in slopes is significant at the 3$\sigma$ level indicating a possible dependence on period. This could be because bolometric corrections depend not only on metallicities but also on gravity and temperature (or colour) \citep{kudritzki2008} where the latter may contribute to the dependence on the period through period-colour relations. We emphasize that the bolometric corrections of \citet{lejeune1998} essentially come from the theoretical SEDs where the full coverage of the atmospheric parameters is ensured by combining the synthetic spectra from the \citet{kurucz1970} atmospheric models, supplemented with M giants spectra from \citet{fluks1994} and \citet{bessell1989, bessell1991} and spectra of M dwarfs from \citet{allard1995} at low temperatures. However, a detailed investigation of the impact of different model atmospheres and the adopted bolometric corrections on the $PL$ relations is beyond the scope of the present study.

\begin{table}
\caption{$PLZ$ relations for BL~Her models of the mathematical form $M_\lambda=a+b\log(P)+c\mathrm{[Fe/H]}$ for different wavelengths using different convective parameter sets.}
\centering
\scalebox{0.9}{
\begin{tabular}{c c c c c c}
\hline\hline
Band & $a$ & $b$ & $c$ & $\sigma$ & $N$\\
\hline \hline
\multicolumn{6}{c}{Convection set A}\\
\hline
U & 0.438 $\pm$ 0.017 & -0.998 $\pm$ 0.041 & 0.234 $\pm$ 0.009 & 0.357 & 3266\\
B & 0.276 $\pm$ 0.016 & -1.221 $\pm$ 0.04 & 0.082 $\pm$ 0.009 & 0.347 & 3266\\
V & -0.134 $\pm$ 0.013 & -1.62 $\pm$ 0.032 & 0.006 $\pm$ 0.007 & 0.284 & 3266\\
R & -0.372 $\pm$ 0.012 & -1.848 $\pm$ 0.028 & -0.007 $\pm$ 0.006 & 0.248 & 3266\\
I & -0.593 $\pm$ 0.01 & -2.043 $\pm$ 0.025 & -0.001 $\pm$ 0.006 & 0.219 & 3266\\
J & -0.908 $\pm$ 0.009 & -2.306 $\pm$ 0.021 & 0.005 $\pm$ 0.005 & 0.186 & 3266\\
H & -1.154 $\pm$ 0.007 & -2.58 $\pm$ 0.018 & 0.015 $\pm$ 0.004 & 0.156 & 3266\\
K & -1.107 $\pm$ 0.008 & -2.539 $\pm$ 0.018 & 0.015 $\pm$ 0.004 & 0.16 & 3266\\
L & -1.212 $\pm$ 0.007 & -2.611 $\pm$ 0.018 & 0.017 $\pm$ 0.004 & 0.154 & 3266\\
L' & -1.214 $\pm$ 0.007 & -2.611 $\pm$ 0.018 & 0.019 $\pm$ 0.004 & 0.155 & 3266\\
M & -1.441 $\pm$ 0.007 & -2.833 $\pm$ 0.016 & 0.047 $\pm$ 0.004 & 0.14 & 3266\\
Bolometric & -0.131 $\pm$ 0.012 & -1.83 $\pm$ 0.029 & 0.047 $\pm$ 0.006 & 0.251 & 3266\\
\hline
\multicolumn{6}{c}{Convection set B}\\
\hline
U & 0.454 $\pm$ 0.017 & -0.706 $\pm$ 0.042 & 0.234 $\pm$ 0.01 & 0.315 & 2260\\
B & 0.284 $\pm$ 0.016 & -0.93 $\pm$ 0.04 & 0.073 $\pm$ 0.009 & 0.307 & 2260\\
V & -0.138 $\pm$ 0.014 & -1.372 $\pm$ 0.034 & -0.004 $\pm$ 0.008 & 0.256 & 2260\\
R & -0.381 $\pm$ 0.012 & -1.627 $\pm$ 0.03 & -0.016 $\pm$ 0.007 & 0.227 & 2260\\
I & -0.607 $\pm$ 0.011 & -1.844 $\pm$ 0.027 & -0.008 $\pm$ 0.006 & 0.203 & 2260\\
J & -0.931 $\pm$ 0.009 & -2.129 $\pm$ 0.023 & -0.003 $\pm$ 0.005 & 0.177 & 2260\\
H & -1.184 $\pm$ 0.008 & -2.433 $\pm$ 0.02 & 0.008 $\pm$ 0.005 & 0.154 & 2260\\
K & -1.136 $\pm$ 0.008 & -2.387 $\pm$ 0.021 & 0.008 $\pm$ 0.005 & 0.157 & 2260\\
L & -1.242 $\pm$ 0.008 & -2.466 $\pm$ 0.02 & 0.011 $\pm$ 0.005 & 0.153 & 2260\\
L' & -1.244 $\pm$ 0.008 & -2.467 $\pm$ 0.02 & 0.012 $\pm$ 0.005 & 0.153 & 2260\\
M & -1.475 $\pm$ 0.008 & -2.715 $\pm$ 0.019 & 0.043 $\pm$ 0.004 & 0.142 & 2260\\
Bolometric & -0.141 $\pm$ 0.012 & -1.599 $\pm$ 0.03 & 0.037 $\pm$ 0.007 & 0.229 & 2260\\
\hline
\multicolumn{6}{c}{Convection set C}\\
\hline
U & 0.589 $\pm$ 0.02 & -0.646 $\pm$ 0.047 & 0.271 $\pm$ 0.011 & 0.387 & 2632\\
B & 0.423 $\pm$ 0.018 & -1.014 $\pm$ 0.043 & 0.087 $\pm$ 0.01 & 0.354 & 2632\\
V & -0.029 $\pm$ 0.015 & -1.488 $\pm$ 0.035 & 0.002 $\pm$ 0.008 & 0.288 & 2632\\
R & -0.294 $\pm$ 0.013 & -1.729 $\pm$ 0.031 & -0.012 $\pm$ 0.007 & 0.253 & 2632\\
I & -0.537 $\pm$ 0.012 & -1.929 $\pm$ 0.027 & -0.003 $\pm$ 0.007 & 0.226 & 2632\\
J & -0.878 $\pm$ 0.01 & -2.238 $\pm$ 0.023 & -0.001 $\pm$ 0.006 & 0.19 & 2632\\
H & -1.15 $\pm$ 0.008 & -2.539 $\pm$ 0.019 & 0.011 $\pm$ 0.005 & 0.16 & 2632\\
K & -1.1 $\pm$ 0.008 & -2.492 $\pm$ 0.02 & 0.011 $\pm$ 0.005 & 0.164 & 2632\\
L & -1.21 $\pm$ 0.008 & -2.572 $\pm$ 0.019 & 0.015 $\pm$ 0.005 & 0.158 & 2632\\
L' & -1.211 $\pm$ 0.008 & -2.574 $\pm$ 0.019 & 0.016 $\pm$ 0.005 & 0.158 & 2632\\
M & -1.456 $\pm$ 0.007 & -2.803 $\pm$ 0.017 & 0.054 $\pm$ 0.004 & 0.143 & 2632\\
Bolometric & -0.047 $\pm$ 0.013 & -1.729 $\pm$ 0.031 & 0.043 $\pm$ 0.007 & 0.254 & 2632\\
\hline
\multicolumn{6}{c}{Convection set D}\\
\hline
U & 0.588 $\pm$ 0.02 & -0.527 $\pm$ 0.048 & 0.273 $\pm$ 0.012 & 0.364 & 2122\\
B & 0.411 $\pm$ 0.018 & -0.855 $\pm$ 0.044 & 0.086 $\pm$ 0.011 & 0.334 & 2122\\
V & -0.043 $\pm$ 0.015 & -1.337 $\pm$ 0.036 & -0.0 $\pm$ 0.009 & 0.275 & 2122\\
R & -0.307 $\pm$ 0.013 & -1.593 $\pm$ 0.032 & -0.013 $\pm$ 0.008 & 0.243 & 2122\\
I & -0.549 $\pm$ 0.012 & -1.808 $\pm$ 0.029 & -0.004 $\pm$ 0.007 & 0.218 & 2122\\
J & -0.892 $\pm$ 0.01 & -2.122 $\pm$ 0.025 & -0.002 $\pm$ 0.006 & 0.187 & 2122\\
H & -1.164 $\pm$ 0.009 & -2.438 $\pm$ 0.021 & 0.011 $\pm$ 0.005 & 0.161 & 2122\\
K & -1.114 $\pm$ 0.009 & -2.389 $\pm$ 0.022 & 0.01 $\pm$ 0.005 & 0.165 & 2122\\
L & -1.224 $\pm$ 0.009 & -2.472 $\pm$ 0.021 & 0.014 $\pm$ 0.005 & 0.16 & 2122\\
L' & -1.225 $\pm$ 0.009 & -2.474 $\pm$ 0.021 & 0.016 $\pm$ 0.005 & 0.16 & 2122\\
M & -1.468 $\pm$ 0.008 & -2.723 $\pm$ 0.019 & 0.054 $\pm$ 0.005 & 0.147 & 2122\\
Bolometric & -0.062 $\pm$ 0.013 & -1.582 $\pm$ 0.032 & 0.04 $\pm$ 0.008 & 0.244 & 2122\\
\hline
\multicolumn{6}{c}{Theoretical relations from \citet{criscienzo2007}$^{\dagger}$}\\
\hline
I & -0.26$\pm$0.19 & -2.10$\pm$0.06 & 0.04$\pm$0.01 & - & -\\
J & -0.64$\pm$0.13 & -2.29$\pm$0.04 & 0.04$\pm$0.01 & - & -\\
H & -0.95$\pm$0.06 & -2.34$\pm$0.02 & 0.06$\pm$0.01 & - & -\\
K & -0.97$\pm$0.06 & -2.38$\pm$0.02 & 0.06$\pm$0.01 & - & -\\
\hline
\end{tabular}}
\begin{tablenotes}
	\small
	\item $^{\dagger}$ For BL~Her models with $-2.62\; \mathrm{dex} \leq \mathrm{[Fe/H]} \leq -0.66\; \mathrm{dex}$ and $0.8 < \mathrm{P (days)} < 8$  
\end{tablenotes}
\label{tab:PLZ}
\end{table}

\begin{figure*}
\centering
\includegraphics[scale = 1]{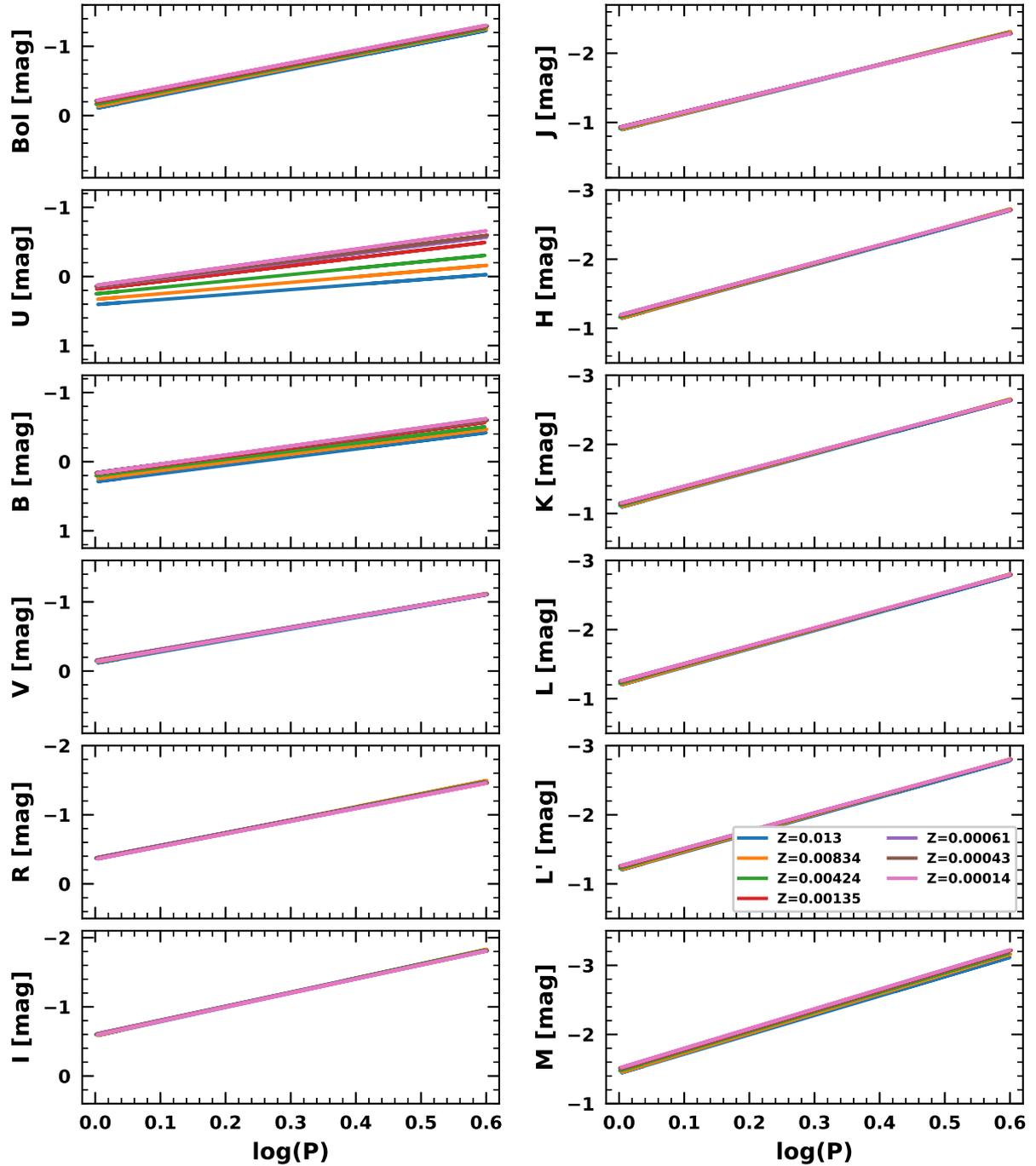}
\caption{$PL$ relations of the BL~Her models with different chemical compositions across different wavelengths for the convective parameter set~A. The y-scale is same (2.5~mag) in each panel for a relative comparison. The other convection parameter sets (B, C and D) show similar $PL$ relations as a function of metallicity and wavelength.}
\label{fig:PL_Zlambda}
\end{figure*}

Fig.~\ref{fig:PL_Zlambda} presents the $PL$ relations of the BL~Her models with different chemical compositions across different wavelengths for the convective parameter set~A. The other convection parameter sets (B, C and D) show similar $PL$ relations as a function of metallicity and wavelength. From Fig.~\ref{fig:PL_Zlambda}, we find that the different chemical compositions lead to statistically similar slopes of $PL$ relations. 

\subsection{Period-Wesenheit relations}

Wesenheit indices \citep{madore1982} may be used as pseudo-magnitudes but with the added advantage that they are minimally affected by the uncertainties related to reddening corrections. A few empirical studies provide $PW$ relations instead of $PL$ relations \citep[see example,][]{bhardwaj2017b}. To facilitate the ease of comparison with empirical results, we provide theoretical $PW$ relations for our models. For the magnitudes ($m_{\lambda_1}$ and $m_{\lambda_2}$) in two bands ($\lambda_1$ and $\lambda_2$), the Wesenheit index may be defined as \citep{inno2013, bhardwaj2016a}:
\begin{equation}
W(\lambda_2, \lambda_1)=m_{\lambda_1} - \left [\frac{A(\lambda_1)}{E(m_{\lambda_2}-m_{\lambda_1})} \right ] \times (m_{\lambda_2}-m_{\lambda_1}),
\end{equation}
where $\lambda_1 > \lambda_2$ and $A(\lambda_1)/E(m_{\lambda_2}-m_{\lambda_1})$ is the total-to-selective extinction for the given filters using a particular reddening law. We adopt the reddening law from \citet{cardelli1989} and assume $R_V = A(V)/(A(B)-A(V)) = 3.23$. For our study, we combine five optical-NIR ($VIJHK$) mean magnitudes to obtain 10 Wesenheit indices using the selective absorption ratios \citep{inno2013}: $A_I/A_V=0.61$, $A_J/A_V=0.29$, $A_H/A_V=0.18$ and $A_K/A_V=0.12$. We derive the corresponding $PW$ relations of the mathematical form $W=a\log(P)+b$ for the BL~Her models. The statistical comparison of the slopes from the theoretical NIR and optical-NIR $PW$ relations of the BL~Her models using different convection parameter sets and those from previous literature is provided in Table~\ref{tab:PW}. Similar to the $PL$ slopes, the BL~Her models with radiative cooling (sets B and D) exhibit statistically similar $PW$ slopes across all 10 Wesenheit indices, while the models without radiative cooling (sets A and C) present larger differences for $W_{VI}$ and $W_{HK}$ indices. The theoretical $PW$ relations from the models are consistent with the empirical relations for BL~Her stars in the LMC and the SMC \citep{matsunaga2009, matsunaga2011, groenewegen2017b}. We note here that the differences between the theoretical $PW$ relations from the models are smaller than empirical uncertainties from the data.

\begin{table*}
\caption{Comparison of the slopes of the NIR and optical-NIR $PW$ relations for BL~Her stars of the mathematical form $W=a\log(P)+b$. $N$ is the total number of stars and the number in parentheses indicates the number of outliers. |$T$| represents the observed value of the $t$-statistic, and $p(t)$ gives the probability of acceptance of the null hypothesis (equal slopes). The bold-faced entries indicate that the null hypothesis of the equivalent $PL$ slopes can be rejected.}
\centering
\scalebox{0.8}{
\begin{tabular}{c c c c c c c c c c c c}
\hline\hline
Band & Source & $a$ & $b$ & $\sigma$ & $N$ & Reference${^\ddagger}$ & \makecell[c]{Theoretical/ \\ Empirical} & \multicolumn{4}{c}{(|$T$|, $p(t)$) w.r.t.}\\
& & & & & & & & Set A & Set B & Set C & Set D\\
\hline \hline
$\rm{W_{VI}}$ & $\rm{Z_{all}}$ (Set A) &-2.705$\pm$0.017&-1.292$\pm$0.006&0.147&3266& TW & Theoretical & ... & ... & ... & ...\\
$\rm{W_{VI}}$ & $\rm{Z_{all}}$ (Set B) &-2.582$\pm$0.019&-1.32$\pm$0.006&0.147&2260& TW & Theoretical & \textbf{(4.877,0.0)}& ... & ... & ...\\
$\rm{W_{VI}}$ & $\rm{Z_{all}}$ (Set C) &-2.622$\pm$0.018&-1.314$\pm$0.006&0.153&2632& TW & Theoretical & \textbf{(3.376,0.0)} & (1.528,0.063)  & ... & ...\\
$\rm{W_{VI}}$ & $\rm{Z_{all}}$ (Set D) &-2.542$\pm$0.02&-1.325$\pm$0.007&0.155&2122& TW & Theoretical & \textbf{(6.273,0.0)} & (1.417,0.078) & \textbf{(2.945,0.002)} &  ...\\

$\rm{W_{VI}}$ & LMC & -2.598$\pm$0.094 & 16.597$\pm$0.017 (@0.3)$^{*}$ & 0.10 & 55 & M09 & Empirical &(1.125,0.13) & (0.169,0.433) & (0.253,0.4) & (0.579,0.281)\\
$\rm{W_{VI}}$ & LMC & -2.576$\pm$0.080 & 17.359$\pm$0.022 & 0.089 & 55(6) & G17 & Empirical &(1.584,0.057) & (0.07,0.472) & (0.564,0.286) & (0.408,0.342)\\
$\rm{W_{VI}}$ & @LMC$^{\dagger}$ & -2.669$\pm$0.137 & 17.347$\pm$0.038 & 0.170 & 74(4) & G17 & Empirical &(0.263,0.396) & (0.631,0.264) & (0.338,0.368) & (0.915,0.18)\\
$\rm{W_{VI}}$ & SMC & -2.421$\pm$0.479 & 16.832$\pm$0.069 (@0.3)$^{*}$ & 0.26 & 17 & M11 & Empirical &(0.593,0.277) & (0.335,0.369) & (0.42,0.337) & (0.253,0.4)\\
$\rm{W_{VI}}$ & SMC & -2.429$\pm$0.480 & 17.558$\pm$0.134 & 0.241 & 17(0) & G17 & Empirical &(0.575,0.283) & (0.318,0.375) & (0.402,0.344) & (0.236,0.407)\\
\hline
$\rm{W_{VJ}}$ & $\rm{Z_{all}}$ (Set A) &-2.584$\pm$0.018&-1.23$\pm$0.006&0.156&3266& TW & Theoretical & ... & ... & ... & ...\\
$\rm{W_{VJ}}$ & $\rm{Z_{all}}$ (Set B) &-2.441$\pm$0.02&-1.254$\pm$0.007&0.154&2260& TW & Theoretical & \textbf{(5.349,0.0)}& ... & ... & ...\\
$\rm{W_{VJ}}$ & $\rm{Z_{all}}$ (Set C) &-2.547$\pm$0.019&-1.224$\pm$0.007&0.16&2632& TW & Theoretical & (1.422,0.078) & \textbf{(3.831,0.0)}  & ... & ...\\
$\rm{W_{VJ}}$ & $\rm{Z_{all}}$ (Set D) &-2.444$\pm$0.021&-1.238$\pm$0.007&0.161&2122& TW & Theoretical & \textbf{(5.11,0.0)} & (0.127,0.449) & \textbf{(3.631,0.0)} &  ...\\
\hline
$\rm{W_{VH}}$ & $\rm{Z_{all}}$ (Set A) &-2.78$\pm$0.016&-1.396$\pm$0.005&0.142&3266& TW & Theoretical & ... & ... & ... & ...\\
$\rm{W_{VH}}$ & $\rm{Z_{all}}$ (Set B) &-2.662$\pm$0.019&-1.424$\pm$0.006&0.143&2260& TW & Theoretical & \textbf{(4.801,0.0)}& ... & ... & ...\\
$\rm{W_{VH}}$ & $\rm{Z_{all}}$ (Set C) &-2.759$\pm$0.017&-1.411$\pm$0.006&0.144&2632& TW & Theoretical & (0.907,0.182) & \textbf{(3.82,0.0)}  & ... & ...\\
$\rm{W_{VH}}$ & $\rm{Z_{all}}$ (Set D) &-2.673$\pm$0.019&-1.424$\pm$0.007&0.148&2122& TW & Theoretical & \textbf{(4.296,0.0)} & (0.411,0.341) & \textbf{(3.343,0.0)} &  ...\\
\hline
$\rm{W_{VK}}$ & $\rm{Z_{all}}$ (Set A) &-2.647$\pm$0.017&-1.251$\pm$0.006&0.15&3266& TW & Theoretical & ... & ... & ... & ...\\
$\rm{W_{VK}}$ & $\rm{Z_{all}}$ (Set B) &-2.514$\pm$0.02&-1.276$\pm$0.006&0.15&2260& TW & Theoretical & \textbf{(5.132,0.0)}& ... & ... & ...\\
$\rm{W_{VK}}$ & $\rm{Z_{all}}$ (Set C) &-2.613$\pm$0.018&-1.253$\pm$0.006&0.153&2632& TW & Theoretical & (1.378,0.084) & \textbf{(3.68,0.0)}  & ... & ...\\
$\rm{W_{VK}}$ & $\rm{Z_{all}}$ (Set D) &-2.519$\pm$0.02&-1.266$\pm$0.007&0.156&2122& TW & Theoretical & \textbf{(4.854,0.0)} & (0.181,0.428) & \textbf{(3.436,0.0)} &  ...\\
\hline
$\rm{W_{IJ}}$ & $\rm{Z_{all}}$ (Set A) &-2.541$\pm$0.018&-1.209$\pm$0.006&0.161&3266& TW & Theoretical & ... & ... & ... & ...\\
$\rm{W_{IJ}}$ & $\rm{Z_{all}}$ (Set B) &-2.391$\pm$0.021&-1.231$\pm$0.007&0.158&2260& TW & Theoretical & \textbf{(5.471,0.0)}& ... & ... & ...\\
$\rm{W_{IJ}}$ & $\rm{Z_{all}}$ (Set C) &-2.521$\pm$0.019&-1.193$\pm$0.007&0.163&2632& TW & Theoretical & (0.758,0.224) & \textbf{(4.587,0.0)}  & ... & ...\\
$\rm{W_{IJ}}$ & $\rm{Z_{all}}$ (Set D) &-2.41$\pm$0.021&-1.208$\pm$0.007&0.165&2122& TW & Theoretical & \textbf{(4.685,0.0)} & (0.648,0.259) & \textbf{(3.843,0.0)} &  ...\\
\hline
$\rm{W_{IH}}$ & $\rm{Z_{all}}$ (Set A) &-2.792$\pm$0.016&-1.412$\pm$0.005&0.142&3266& TW & Theoretical & ... & ... & ... & ...\\
$\rm{W_{IH}}$ & $\rm{Z_{all}}$ (Set B) &-2.674$\pm$0.019&-1.44$\pm$0.006&0.143&2260& TW & Theoretical & \textbf{(4.779,0.0)}& ... & ... & ...\\
$\rm{W_{IH}}$ & $\rm{Z_{all}}$ (Set C) &-2.78$\pm$0.017&-1.426$\pm$0.006&0.143&2632& TW & Theoretical & (0.479,0.316) & \textbf{(4.203,0.0)}  & ... & ...\\
$\rm{W_{IH}}$ & $\rm{Z_{all}}$ (Set D) &-2.693$\pm$0.019&-1.439$\pm$0.007&0.147&2122& TW & Theoretical & \textbf{(3.949,0.0)} & (0.721,0.235) & \textbf{(3.404,0.0)}&  ...\\
\hline
$\rm{W_{IK}}$ & $\rm{Z_{all}}$ (Set A) &-2.645$\pm$0.017&-1.251$\pm$0.006&0.151&3266& TW & Theoretical & ... & ... & ... & ...\\
$\rm{W_{IK}}$ & $\rm{Z_{all}}$ (Set B) &-2.511$\pm$0.02&-1.275$\pm$0.006&0.15&2260& TW & Theoretical & \textbf{(5.143,0.0)}& ... & ... & ...\\
$\rm{W_{IK}}$ & $\rm{Z_{all}}$ (Set C) &-2.615$\pm$0.018&-1.251$\pm$0.006&0.153&2632& TW & Theoretical & (1.18,0.119) & \textbf{(3.88,0.0)}  & ... & ...\\
$\rm{W_{IK}}$ & $\rm{Z_{all}}$ (Set D) &-2.521$\pm$0.02&-1.264$\pm$0.007&0.156&2122& TW & Theoretical & \textbf{(4.706,0.0)} & (0.334,0.369) & \textbf{(3.477,0.0)} &  ...\\
\hline
$\rm{W_{JH}}$ & $\rm{Z_{all}}$ (Set A) &-3.006$\pm$0.015&-1.587$\pm$0.005&0.137&3266& TW & Theoretical & ... & ... & ... & ...\\
$\rm{W_{JH}}$ & $\rm{Z_{all}}$ (Set B) &-2.917$\pm$0.018&-1.621$\pm$0.006&0.139&2260& TW & Theoretical & \textbf{(3.762,0.0)}& ... & ... & ...\\
$\rm{W_{JH}}$ & $\rm{Z_{all}}$ (Set C) &-3.003$\pm$0.016&-1.627$\pm$0.006&0.136&2632& TW & Theoretical & (0.146,0.442) & \textbf{(3.544,0.0)}  & ... & ...\\
$\rm{W_{JH}}$ & $\rm{Z_{all}}$ (Set D) &-2.936$\pm$0.018&-1.639$\pm$0.006&0.142&2122& TW & Theoretical & \textbf{(2.919,0.002)} & (0.753,0.226) & \textbf{(2.722,0.003)} &  ...\\
\hline
$\rm{W_{JK}}$ & $\rm{Z_{all}}$ (Set A) &-2.684$\pm$0.017&-1.268$\pm$0.006&0.148&3266& TW & Theoretical & ... & ... & ... & ...\\
$\rm{W_{JK}}$ & $\rm{Z_{all}}$ (Set B) &-2.557$\pm$0.019&-1.294$\pm$0.006&0.147&2260& TW & Theoretical & \textbf{(4.991,0.0)}& ... & ... & ...\\
$\rm{W_{JK}}$ & $\rm{Z_{all}}$ (Set C) &-2.652$\pm$0.018&-1.275$\pm$0.006&0.15&2632& TW & Theoretical & (1.329,0.092) & \textbf{(3.597,0.0)}  & ... & ...\\
$\rm{W_{JK}}$ & $\rm{Z_{all}}$ (Set D) &-2.563$\pm$0.02&-1.287$\pm$0.007&0.153&2122& TW & Theoretical & \textbf{(4.682,0.0)} & (0.218,0.414) & \textbf{(3.32,0.0)} &  ...\\
\hline
$\rm{W_{HK}}$ & $\rm{Z_{all}}$ (Set A) &-2.448$\pm$0.019&-1.035$\pm$0.006&0.168&3266& TW & Theoretical & ... & ... & ... & ...\\
$\rm{W_{HK}}$ & $\rm{Z_{all}}$ (Set B) &-2.293$\pm$0.021&-1.054$\pm$0.007&0.163&2260& TW & Theoretical & \textbf{(5.421,0.0)}& ... & ... & ...\\
$\rm{W_{HK}}$ & $\rm{Z_{all}}$ (Set C) &-2.394$\pm$0.02&-1.016$\pm$0.007&0.172&2632& TW & Theoretical & \textbf{(1.932,0.027)} & \textbf{(3.407,0.0)}  & ... & ...\\
$\rm{W_{HK}}$ & $\rm{Z_{all}}$ (Set D) &-2.289$\pm$0.022&-1.029$\pm$0.008&0.171&2122& TW & Theoretical & \textbf{(5.441,0.0)} & (0.124,0.451) & \textbf{(3.468,0.0)} &  ...\\
\hline
\end{tabular}}
\begin{tablenotes}
	\small
	\item ${^\ddagger}$ TW=This work; M09=\citet{matsunaga2009}; M11=\citet{matsunaga2011}; G17=\citet{groenewegen2017b}
	\item $^{*}$ Zero point at $\log(P)=0.3$    
	\item $^{\dagger}$ The stars in the LMC plus the stars in the SMC placed at the distance of the LMC by a shift of 0.432 mag   
\end{tablenotes}
\label{tab:PW}
\end{table*}

\section{Period-radius relations}
\label{sec:PR}

The mean radius obtained from averaging the radius of the BL~Her model over a pulsation cycle may be used for deriving theoretical $PR$ relations for BL~Her models of the mathematical form \citep{burki1986, marconi2007}:

\begin{equation}
\log(R/R_{\odot})=\alpha \log(P)+\beta
\end{equation}

\subsection{Effect of convection parameters on $PR$ relations}
The $PR$ relations for BL~Her models for different chemical compositions using different convective parameter sets are summarised in Table~\ref{tab:PR_differentZ}. The slopes and intercepts of the $PR$ relations obtained using different chemical compositions are found to be similar. 

\begin{table}
\caption{Period-radius relations for BL~Her models of the mathematical form $\log(R/R_{\odot})=\alpha \log(P)+\beta$ for different chemical compositions using different convective parameter sets.}
\centering
\begin{tabular}{c c c c c}
\hline\hline
Source & $\alpha$ & $\beta$ & $\sigma$ & $N$\\
\hline \hline
\multicolumn{5}{c}{Convection set A}\\
\hline
Z=0.00014 & 0.574$\pm$0.009&0.892$\pm$0.003&0.029&434\\
Z=0.00043 & 0.572$\pm$0.009&0.891$\pm$0.003&0.029&432\\
Z=0.00061 & 0.574$\pm$0.009&0.89$\pm$0.003&0.029&431\\
Z=0.00135 & 0.576$\pm$0.009&0.887$\pm$0.003&0.029&437\\
Z=0.00424 & 0.584$\pm$0.009&0.882$\pm$0.003&0.029&466\\
Z=0.00834 & 0.594$\pm$0.008&0.876$\pm$0.003&0.028&515\\
Z=0.01300 & 0.588$\pm$0.007&0.875$\pm$0.003&0.027&551\\
\hline
\multicolumn{5}{c}{Convection set B}\\
\hline
Z=0.00014 & 0.534$\pm$0.011&0.902$\pm$0.003&0.03&302\\
Z=0.00043 & 0.536$\pm$0.011&0.9$\pm$0.003&0.029&299\\
Z=0.00061 & 0.535$\pm$0.01&0.9$\pm$0.003&0.029&302\\
Z=0.00135 & 0.535$\pm$0.01&0.896$\pm$0.003&0.029&314\\
Z=0.00424 & 0.545$\pm$0.01&0.891$\pm$0.003&0.028&319\\
Z=0.00834 & 0.57$\pm$0.01&0.882$\pm$0.003&0.029&340\\
Z=0.01300 & 0.572$\pm$0.009&0.88$\pm$0.003&0.029&384\\
\hline
\multicolumn{5}{c}{Convection set C}\\
\hline
Z=0.00014 & 0.56$\pm$0.011&0.897$\pm$0.003&0.029&314\\
Z=0.00043 & 0.563$\pm$0.011&0.895$\pm$0.003&0.029&321\\
Z=0.00061 & 0.565$\pm$0.01&0.894$\pm$0.003&0.029&324\\
Z=0.00135 & 0.572$\pm$0.01&0.891$\pm$0.003&0.028&333\\
Z=0.00424 & 0.576$\pm$0.009&0.885$\pm$0.003&0.028&380\\
Z=0.00834 & 0.598$\pm$0.008&0.876$\pm$0.003&0.027&444\\
Z=0.01300 & 0.599$\pm$0.008&0.872$\pm$0.003&0.028&516\\
\hline
\multicolumn{5}{c}{Convection set D}\\
\hline
Z=0.00014 & 0.53$\pm$0.011&0.902$\pm$0.003&0.03&262\\
Z=0.00043 & 0.534$\pm$0.011&0.901$\pm$0.004&0.031&275\\
Z=0.00061 & 0.541$\pm$0.01&0.898$\pm$0.003&0.029&280\\
Z=0.00135 & 0.543$\pm$0.01&0.894$\pm$0.003&0.029&277\\
Z=0.00424 & 0.559$\pm$0.01&0.886$\pm$0.003&0.028&289\\
Z=0.00834 & 0.57$\pm$0.01&0.882$\pm$0.004&0.029&336\\
Z=0.01300 & 0.59$\pm$0.009&0.874$\pm$0.003&0.028&403\\
\hline
\end{tabular}
\label{tab:PR_differentZ}
\end{table}

\begin{figure}
\centering
\includegraphics[scale = 1]{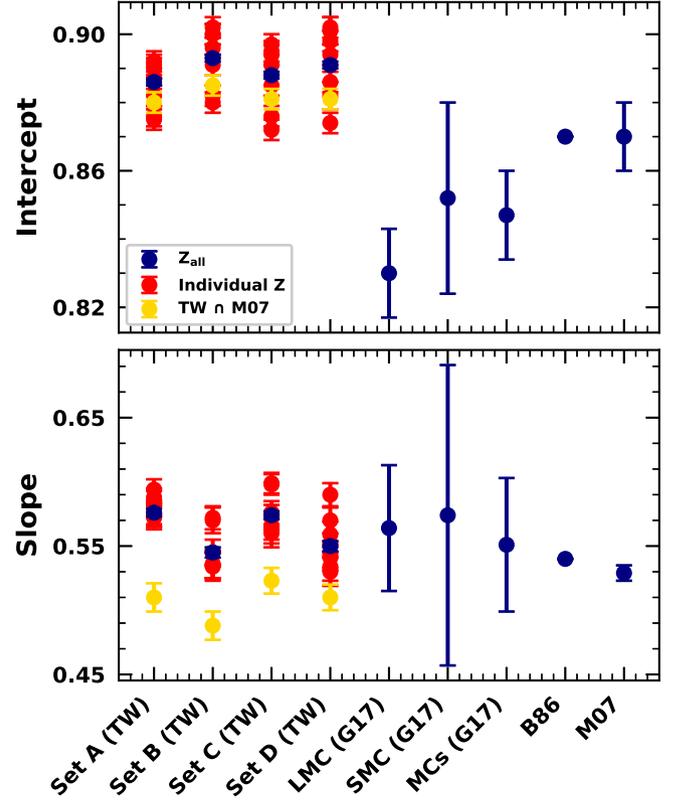}
\caption{A comparison of the slopes and intercepts of the $PR$ relations for the BL~Her stars obtained from this work (TW) using four different convective parameter sets with those obtained from previous literature. The red dots refer to the slopes and intercepts of the $PR$ relations obtained for the different chemical compositions individually, while the blue dots refer to the results obtained from considering the entire range of chemical compositions ($Z=0.00014-Z=0.013$) together. The yellow dots represent the subset of the model grid of this work with the same parameter space as that of \citet{marconi2007}. G17, B86 and M07 refer to \citet{groenewegen2017b}, \citet{burki1986} and \citet{marconi2007}, respectively.}
\label{fig:PR}
\end{figure}

We use the statistical $t$-test (Eq.~\ref{eq:ttest}) to compare the slopes from the theoretical $PR$ relations across different convection parameter sets with those obtained from previous studies. The results from this test are presented in Table~\ref{tab:PR}. Models with sets A and C have statistically similar $PR$ slopes while those with sets B and D show similar $PR$ slopes. We also find the slopes from the theoretical $PR$ relations of the BL~Her models to be similar with those from the empirical results for the LMC and the SMC from \citet{groenewegen2017b}. Fig.\ref{fig:PR} presents a comparison of the slopes and intercepts of the $PR$ relations for the BL~Her stars obtained from this work using four different convective parameter sets with those obtained from previous literature. We also use a subset of our BL~Her models with the same input parameter space as that of \citet{marconi2007} to compare the theoretical $PR$ relations; the results are displayed in Fig.\ref{fig:PR}.

\begin{table*}
\caption{Comparison of the slopes of the $PR$ relations for BL~Her stars of the mathematical form $\log(R/R_{\odot})=\alpha \log(P)+\beta$. $N$ is the total number of stars and the number in parentheses indicates the number of outliers. |$T$| represents the observed value of the $t$-statistic, and $p(t)$ gives the probability of acceptance of the null hypothesis (equal slopes). The bold-faced entries indicate that the null hypothesis of the equivalent $PR$ slopes can be rejected.}
\scalebox{0.8}{
\begin{tabular}{c c c c c c c c c c c}
\hline\hline
Source & $\alpha$ & $\beta$ & $\sigma$ & $N$ & Reference${^\ddagger}$ & \makecell[c]{Theoretical/ \\ Empirical} & \multicolumn{4}{c}{(|$T$|, $p(t)$) w.r.t.}\\
& & & & & & & Set A & Set B & Set C & Set D\\
\hline \hline
$\rm{Z_{all}}$ (Set A) &0.576$\pm$0.003&0.886$\pm$0.001&0.029&3266& TW & Theoretical & ... & ... & ... & ...\\
$\rm{Z_{all}}$ (Set B) &0.545$\pm$0.004&0.893$\pm$0.001&0.029&2260& TW & Theoretical & \textbf{(6.278,0.0)}& ... & ... & ...\\
$\rm{Z_{all}}$ (Set C) &0.574$\pm$0.003&0.888$\pm$0.001&0.029&2632& TW & Theoretical & (0.588,0.278) & \textbf{(5.595,0.0)} & ... & ...\\
$\rm{Z_{all}}$ (Set D) &0.55$\pm$0.004&0.891$\pm$0.001&0.029&2122& TW & Theoretical & \textbf{(5.17,0.0)} & (1.046,0.148) & \textbf{(4.509,0.0)} & ...\\
LMC & 0.564$\pm$0.049 & 0.830$\pm$0.013 & 0.047 & 57(4)& G17 & Empirical & (0.253,0.4) & (0.39,0.348) & (0.197,0.422) & (0.275,0.392)\\
SMC & 0.574$\pm$0.117 & 0.852$\pm$0.028 & 0.056 & 17(0)& G17 & Empirical & (0.021,0.492) & (0.249,0.402) & (0.003,0.499) & (0.201,0.42)\\
MCs & 0.551$\pm$0.052 & 0.847$\pm$0.013 & 0.058 & 76(2)& G17 & Empirical & (0.488,0.313) & (0.119,0.453) & (0.435,0.332) & (0.01,0.496)\\
Galactic T2Cs & 0.54 & 0.87 & - & - & B86$^*$ & Empirical & - & - & - & -\\
$\rm{Z_{all}}$ & 0.529$\pm$0.006 & 0.87$\pm$0.01 & - & - & M07$^*$ & Theoretical & - & - & - & -\\
\hline
\end{tabular}}
\begin{tablenotes}
	\small
	\item ${^\ddagger}$ TW=This work; G17=\citet{groenewegen2017b}; B86=\citet{burki1986}; M07=\citet{marconi2007}
	\item $^{*}$ Data insufficient to determine the value of the $t$-statistic and the associated probability value.
\end{tablenotes}
\label{tab:PR}
\end{table*}

\subsection{Effect of metallicity on $PR$ relations}

To test for the effect of metallicity on the $PR$ relations, we derive $PRZ$ relations of the form:
\begin{equation}
\log(R/R_{\odot})=\alpha + \beta \log(P)+\gamma \mathrm{[Fe/H]}
\end{equation}
for the four different convection parameter sets. We find the following relations for set A:
\begin{equation}
\begin{aligned}
\log(R/R_{\odot})={} & (0.879\pm 0.001)+(0.581\pm 0.003)\log(P)\\
&-(0.006\pm 0.001)\mathrm{[Fe/H]} \: (N=3266; \sigma=0.029),
\end{aligned}
\label{eq:PRZ-A}
\end{equation}
for set B:
\begin{equation}
\begin{aligned}
\log(R/R_{\odot})={}&(0.888\pm 0.002)+(0.548\pm 0.004)\log(P)\\
&-(0.006\pm 0.001)\mathrm{[Fe/H]} \: (N=2260; \sigma=0.029),
\end{aligned}
\end{equation}
for set C:
\begin{equation}
\begin{aligned}
\log(R/R_{\odot})=&{}(0.881\pm 0.001)+(0.579\pm 0.003)\log(P)\\
&-(0.006\pm 0.001)\mathrm{[Fe/H]} \: (N=2632; \sigma=0.028),
\end{aligned}
\end{equation}
and for set D:
\begin{equation}
\begin{aligned}
\log(R/R_{\odot})=&{}(0.886\pm 0.002)+(0.554\pm 0.004)\log(P)\\
&-(0.006\pm 0.001)\mathrm{[Fe/H]} \: (N=2122; \sigma=0.029)
\end{aligned}
\label{eq:PRZ-D}
\end{equation}

From Eqs.~\ref{eq:PRZ-A}-\ref{eq:PRZ-D}, we find the coefficients of the metallicity term to be very small for all four convection parameter sets and thus, we may conclude that there is weak dependence of $PR$ relations on metallicity. This is in agreement with previous empirical results from \citet{burki1986}, \citet{balog1997} and {\citet{groenewegen2017b}}. We also find that the dependency of $PR$ relations on metallicity is identical for all four sets of convection parameters. 

\section{Comparison with RR~Lyrae models}

RR~Lyrae and BL~Her stars are pulsating variables that belong to old stellar populations of similar chemical compositions. Both classes are population~II pulsating stars and offer an important alternative tool to classical Cepheids to calibrate the cosmic distance scale and evaluate the Hubble constant. \citet{majaess2010} presented preliminary evidence of a common $PL$ relation for RR~Lyrae and T2Cs. \citet{ bhardwaj2017b} showed that the $PL$ relation of T2Cs in $K$-band when extended to periods less than 1 day follow the same $PL$ relation as RR~Lyrae stars. They also demonstrated that distances to the GGCs based on T2Cs are consistent with those based on horizontal branch (HB) stars. In the most recent study, \citet{braga2020} derived $PL$ relations of RR~Lyrae and T2Cs in $\omega$~Cen and found empirical evidence that RRab and T2Cs indeed obey the same $JHK_S$ $PL$ relations.

To test this equivalence between RR~Lyrae and T2Cs, we compare the theoretical relations from our BL~Her models with the most recent grid of RR~Lyrae models from \citet{marconi2015}. Table~\ref{tab:RRL_relations} presents the comparison of the $PL$ slopes in $RIJHK_S$ bands for RR~Lyrae models from \citet{marconi2015} with those obtained from the BL~Her models of our work. The RR~Lyrae models exhibit statistically similar $PL$ slopes in the bands, $RIJK_S$ with those from the BL~Her models computed using sets A and C. A possible explanation for the slopes not being statistically similar with BL~Her models computed using sets B and D is that the RR~Lyrae models from \citet{marconi2015} are computed without radiative cooling. Our grid of BL~Her models therefore supports the claim by \citet{braga2020} that the equivalence of the $PL$ relations of RR~Lyrae and T2Cs gives us the opportunity of adopting RRLs+T2Cs together as an alternative to classical Cepheids for the extragalactic distance scale calibration.

From Table~\ref{tab:RRL_relations}, we also find that the $PR$ slope of the RR~Lyrae models from \citet{marconi2015} is statistically similar with those obtained from the BL~Her models in all four sets of convection parameters. This is in support of the result obtained by \citet{marconi2015} where they found similar $PR$ slopes for RR~Lyrae models with theoretical and empirical BL~Her $PR$ slopes from \citet{marconi2007} and \citet{burki1986}, respectively. This therefore suggests tight correlation of evolutionary and pulsational properties of the two classes of evolved low-mass radial variables. A detailed comparison of RR~Lyrae and BL~Her pulsation properties could be useful to probe an evolutionary scenario where BL~Hers are HB stars that have evolved off the HB and are moving up the Asymptotic Giant Branch (AGB). BL~Her stars may then be considered as the evolved component of RR~Lyrae stars \citet{marconi2011, marconi2015}\footnote{In the course of refereeing process of this paper, a new study appeared on ArXiv by Bono et al. (https://ui.adsabs.harvard.edu/abs/2020arXiv200906985B/abstract) discussing in more detail the evolutionary properties of T2Cs}. \citet{marconi2015} cautions that this similarity between BL~Her and RR~Lyrae may not be extended over the entire metallicity range. We note here that \citet{marconi2015} uses the convection formulation outlined in \citet{stellingwerf1982a, stellingwerf1982b} while \textsc{mesa-rsp} uses \citet{kuhfuss1986} turbulent convection theory; it is therefore interesting to obtain similar $PL$ and $PR$ results for RR~Lyrae and BL~Her models using different theories of convection.

\begin{table*}
\caption{Comparison of the slopes of the period-luminosity and period-radius relations of the mathematical forms $M_\lambda=a\log(P)+b$ and $\log(R/R_{\odot})=\alpha \log(P)+\beta$, respectively for RR~Lyrae models from \citet{marconi2015}. $N$ is the total number of models. |$T$| represents the observed value of the $t$-statistic, and $p(t)$ gives the probability of acceptance of the null hypothesis (equal slopes). The bold-faced entries indicate that the null hypothesis of the equivalent $PL$ slopes can be rejected.}
\centering
\scalebox{0.8}{
\begin{tabular}{c c c c c c c c c c c c}
\hline\hline
Band & Source & Slope & Intercept & $\sigma$ & $N$ & Reference${^\ddagger}$ & \makecell[c]{Theoretical/ \\ Empirical} & \multicolumn{4}{c}{(|$T$|, $p(t)$) w.r.t. This work}\\
& & & & & & & & Set A & Set B & Set C & Set D\\
\hline
\multicolumn{12}{c}{Period-luminosity relation}\\
\hline
R & $\rm{Z_{all}}$ &-1.756$\pm$0.077&-0.114$\pm$0.014&0.196&226& M15 & Theoretical & (1.19,0.117) & (1.485,0.069) & (0.208,0.418) & \textbf{(1.882,0.03)}\\
I & $\rm{Z_{all}}$ &-1.973$\pm$0.068&-0.415$\pm$0.013&0.175&226& M15 & Theoretical & (0.966,0.167) & \textbf{(1.709,0.044)} & (0.561,0.287) & \textbf{(2.217,0.013)}\\
J & $\rm{Z_{all}}$ &-2.245$\pm$0.06&-0.778$\pm$0.011&0.155&226& M15 & Theoretical & (0.902,0.184) & \textbf{(1.769,0.039)} & (0.098,0.461) & \textbf{(1.898,0.029)}\\
H & $\rm{Z_{all}}$ &-2.206$\pm$0.118&-1.043$\pm$0.022&0.302&226& M15 & Theoretical & \textbf{(3.056,0.001)} & \textbf{(1.867,0.031)} & \textbf{(2.708,0.003)} & \textbf{(1.889,0.03)}\\
K & $\rm{Z_{all}}$ &-2.514$\pm$0.057&-1.11$\pm$0.011&0.147&226& M15 & Theoretical & (0.24,0.405) & \textbf{(2.149,0.016)} & (0.507,0.306) & \textbf{(2.149,0.016)}\\
\hline
\multicolumn{12}{c}{Period-radius relation}\\
\hline
$-$ & $\rm{Z_{all}}$ &0.557$\pm$0.011&0.871$\pm$0.002&0.029&226& M15 & Theoretical & (1.622,0.052) & (1.015,0.155) & (1.45,0.074) & (0.596,0.276)\\
\hline
\end{tabular}}
\begin{tablenotes}
	\small
	\item ${^\ddagger}$ M15=\citet{marconi2015}   
	\item The $PL$ and $PR$ relations for the RR~Lyrae models have been obtained using Table~3 of \citet{marconi2015}. The $PL$ and $PR$ relations for the BL~Her models computed in this work using the four sets of convection parameters are available in Table~\ref{tab:PL} and Table~\ref{tab:PR}, respectively.
\end{tablenotes}
\label{tab:RRL_relations}
\end{table*}

\section{Summary and Conclusion}
\label{sec:results}

We computed a very fine grid of BL~Her models using the most recent, state-of-the-art stellar pulsation code, the \textsc{mesa-rsp} \citep{paxton2019}. The grid encompasses a wide range of metallicity, mass, luminosity and effective temperature with four different sets of convection parameters, A, B, C and D as outlined in \citet{paxton2019} and Table~\ref{tab:convection}. The metallicity varies from [Fe/H]=-2.0 dex ($Z$=0.00014) to [Fe/H]=0.0 dex ($Z$=0.013). The stellar masses vary from 0.5M$_{\odot}$ to 0.8M$_{\odot}$, while the stellar luminosity varies from 50L$_{\odot}$ to 300L$_{\odot}$, typical for BL~Her stars. Effective temperature is in steps of 50 K inside the instability strip. Non-linear models were computed for 4000 pulsation cycles for periods typical for BL~Her stars, i.e. $1 \leq \mathrm{P (days)} \leq 4$. The non-linear models analysed in this study fulfil the condition of full-amplitude stable pulsations, i.e., amplitude of radius variation $\Delta R$, period $P$ and fractional growth rate $\Gamma$ do not vary by more than 0.01 over the last $\sim$100-cycles of the total 4000-cycle integrations. The total number of BL~Her models accepted are 3266 in set A, 2260 in set B, 2632 in set C and 2122 in set D. We have theoretical lightcurves in multiple wavelengths, $UBVRIJHKLL'M$ from the computed non-linear models.

We obtain theoretical multi-wavelength $PL$, $PW$ and $PR$ relations for these models using the four different sets of convection parameters. We test for the effect of metallicity and convection parameters on the $PL$ and $PR$ relations. We summarise the important results below:

\begin{enumerate}
\item Models computed with sets B and D show statistically similar slopes for $PL$, $PW$ and $PR$ relations while those with sets A and C exhibit similar slopes for most cases. Sets B and D are the models computed with radiative cooling; sets A and C are computed without radiative cooling.
\item Most empirical relations match well with the theoretical $PL$, $PW$ and $PR$ relations derived using our BL~Her models and over all the four sets of convection parameters.
\item An exception to this are the $PL$ relations for BL~Her stars in the LMC where $PL$ slopes of the models with sets B and D seem to match better with empirical relations in the $HK_S$ bands.
\item We find that the $PL$ slopes from the BL~Her models become steeper with increasing wavelengths. The dispersion in the theoretical $PL$ relations for BL~Hers decreases significantly moving from optical to infrared wavelengths and becomes statistically similar for wavelengths longer than $H$-band. We also observe that the models computed using set B have the smallest dispersion in their $PL$ relations in all the bands.
\item For each set of convection parameters, the effect of metallicity is significant in $U$ and $B$-bands, and appreciable for bolometric $PL$ relations. It is much weaker in redder bands, consistent with empirical data and previous studies \citep{matsunaga2006, groenewegen2017b}.
\item There is a weak dependence of the $PR$ relations on metallicity with identical coefficient (0.006 $\pm$ 0.001) for all 4 sets of convection parameters. The inclusion of the metallicity term does not lead to a significant decrease in dispersion for the relations. This is consistent with the empirical evidence from \citet{burki1986}, \citet{balog1997} and {\citet{groenewegen2017b}}.
\item The RR~Lyrae models from \citet{marconi2015} exhibit statistically similar $PL$ relations in the $RIJK_S$ bands with those obtained from BL~Her models computed using sets A and C while the $PR$ slopes from the RR~Lyrae models are statistically similar with the relations from the BL~Her models using all four sets of convection parameters.
\end{enumerate}

However, it is important to note here that both the $PL$ and the $PR$ relations derived in this study are at mean light, which averages out the effect of the pulsation cycle. We also note that the comparison among the models computed using different sets of convection parameters has higher precision than when comparing with empirical $PL$ and $PW$ relations. It would seem that observations are not yet sufficiently precise to distinguish fully among the models, although there is a preference for the models that include radiative cooling. It would be interesting to study the effect of different convective parameter sets on the light curve structures of the BL~Her models during the pulsation cycle, which we plan to investigate in a future project. The large number of models computed in this study ushers in the era of large number statistics in the analysis of theoretical models. 

\section*{Acknowledgements}

The authors thank the referee for useful comments and suggestions that improved the quality of the manuscript. SD acknowledges the INSPIRE Senior Research Fellowship vide Sanction Order No. DST/INSPIRE Fellowship/2016/IF160068 under the INSPIRE Program from the Department of Science \& Technology, Government of India. HPS and SMK thank the Indo-US Science and Technology Forum for funding the Indo-US virtual joint networked centre on ``Theoretical analyses of variable star light curves in the era of large surveys''. SD acknowledges the travel support provided by SERB, Government of India vide file number ITS/2019/004781 to attend the RR Lyrae/Cepheid 2019 Conference ``Frontiers of Classical Pulsators: Theory and Observations'', USA where this work was initiated. RS is supported by the National Science Center, Poland, Sonata BIS project 2018/30/E/ST9/00598. AB acknowledges research grant $\#11850410434$ from the National Natural Science Foundation of China through the Research Fund for International Young Scientists, a China Post-doctoral General Grant, and the Gruber fellowship 2020 grant sponsored by the Gruber Foundation and the International Astronomical Union. The authors acknowledge the use of High Performance Computing facility Pegasus at IUCAA, Pune and the following software used in this project: \textsc{mesa}~r$11701$ \citep{paxton2011,paxton2013,paxton2015,paxton2018,paxton2019}.

\section*{Data Availability}

The data underlying this article are available in the article and in its online supplementary material.

\bibliographystyle{mnras}

\appendix
\section{$PL$ relations using different model atmospheres}
\label{sec:appendix}
\textsc{mesa} provides two sets of pre-processed tables of bolometric corrections \citep{paxton2018}. The results discussed in this work are using the pre-processed table ($\rm{BC_1}$) from \citet{lejeune1998}. To test the impact of the adopted model atmospheres used to transform bolometric light curves to observational bands, we use another pre-processed table ($\rm{BC_2}$) which provides a set of blackbody bolometric corrections for the bands $UBVR_cI_c$ over the range $100 \leq T_{\rm{eff}} \rm{(K)} \leq 50,000$, in steps of 100 K. For the exact same set of models, we compare the slopes of the $PL$ relations in the $V$ and $I$ bands of the mathematical form $\rm{\log_{10} (L_{V/I}/L_{\odot})}=a\log(P)+b$ using the two different sets of bolometric corrections, $\rm{BC_1}$ and $\rm{BC_2}$. We find the $PL_V$ slopes to be $0.657 \pm 0.013$ and $0.68 \pm 0.012$ for 3266 BL~Her models in set A, $0.552 \pm 0.014$ and $0.584 \pm 0.013$ for 2260 models in set B, $0.6 \pm 0.014$ and $0.616 \pm 0.013$ for 2633 models in set C and $0.535 \pm 0.014$ and $0.559 \pm 0.014$ for 2122 models in set D using the two different bolometric correction sets $\rm{BC_1}$ and $\rm{BC_2}$, respectively. The slopes of the $PL$ relations in the $V$ band differ by less than 6\% when a different model atmosphere incorporated in \textsc{mesa} is adopted. For the $I$-band, the $PL$ slopes are $0.825\pm 0.01$ and $0.809\pm 0.01$ in set A, $0.741 \pm 0.011$ and $0.728 \pm 0.011$ in set B, $0.777 \pm 0.011$ and $0.763 \pm 0.011$ in set C and $0.724 \pm 0.011$ and $0.712 \pm 0.011$ in set D using the different bolometric correction sets $\rm{BC_1}$ and $\rm{BC_2}$, respectively. The slopes of the $PL$ relations in the $I$ band therefore differ by less than 2\% when a different model atmosphere is adopted. We find significant variation in $PL$ slopes in $U$ and $B$-band when the two different bolometric sets are adopted. The effect of different model atmospheres on transformations of bolometric light curves to observational bands reduces as we move to longer wavelengths. We note here that the mean luminosity used for these relations have been obtained from averaging the luminosity of the BL~Her model over a pulsation cycle, unlike the mean magnitudes obtained from Fourier fitting as discussed in Section~\ref{sec:PL}.

The reason for adopting the bolometric correction set $\rm{BC_1}$ in our present work is that it provides bolometric corrections for the Johnson-Cousins-Class bands $UBVRIJHKLL'M$ while $\rm{BC_2}$ only provides bolometric corrections for $UBVR_cI_c$. $\rm{BC_1}$ defines the bolometric correction as a function of the stellar photosphere; $T_{\rm{eff}} \rm{(K)}$, $\log (g) \rm{(cm s^{-2})}$, and the metallicity $[M/H]$. Since $\rm{BC_2}$ involves blackbody corrections, there is no $g$ or $[M/H]$ dependence. While a more detailed and quantitative comparison using other model atmospheres is important, it is beyond the scope of this paper and we do not anticipate much difference since we are primarily concerned with $PL$ relations at mean light. Using the two different bolometric correction sets provided by \textsc{mesa} also suggests that the change in $PL$ slopes due to different adopted model transformations is minimal at wavelengths longer than $V$-band. In addition, \citet{paxton2019} demonstrates that the \textsc{mesa-rsp} code produces stable, multi-wavelength light curve models with reasonable comparison with observations for a wide class of radially pulsating stars. Further, \citet{das2020} obtained theoretical period-colour relations for a broad spectrum of variable stars, including BL~Hers which were in broad agreement with observations.

\section{$PLZ$ relations in the low and high-metallicity regimes}

To investigate the dependence of $PL$ relations on metallicity, we separated models in low-metallicity ($Z=0.00014, 0.00043, 0.00061, 0.00135$) and high-metallicity ($Z=0.00424, 0.00834, 0.01300$) regime. The results of $PLZ$ relations for the models in the low and high-metallicity regimes using different convection sets are listed in Tables~\ref{tab:B1} and \ref{tab:B2}, respectively and discussed in Section~\ref{sec:PLZ}.

\begin{table}
\caption{$PLZ$ relations of the mathematical form $M_\lambda=a+b\log(P)+c\mathrm{[Fe/H]}$ for BL~Her models in the low-metallicity regime ($Z=0.00014, 0.00043, 0.00061, 0.00135$) for different wavelengths using different convective parameter sets.}
\centering
\scalebox{0.85}{
\begin{tabular}{c c c c c c}
\hline\hline
Band & $a$ & $b$ & $c$ & $\sigma$ & $N$\\
\hline \hline
\multicolumn{6}{c}{Convection set A}\\
\hline
U & 0.305 $\pm$ 0.036 & -1.217 $\pm$ 0.052 & 0.105 $\pm$ 0.022 & 0.324 & 1734\\
B & 0.207 $\pm$ 0.038 & -1.273 $\pm$ 0.055 & 0.027 $\pm$ 0.023 & 0.34 & 1734\\
V & -0.163 $\pm$ 0.032 & -1.605 $\pm$ 0.046 & -0.01 $\pm$ 0.019 & 0.285 & 1734\\
R & -0.39 $\pm$ 0.028 & -1.827 $\pm$ 0.04 & -0.015 $\pm$ 0.017 & 0.25 & 1734\\
I & -0.611 $\pm$ 0.024 & -2.024 $\pm$ 0.035 & -0.009 $\pm$ 0.015 & 0.22 & 1734\\
J & -0.93 $\pm$ 0.021 & -2.268 $\pm$ 0.03 & -0.003 $\pm$ 0.013 & 0.189 & 1734\\
H & -1.179 $\pm$ 0.018 & -2.543 $\pm$ 0.026 & 0.005 $\pm$ 0.011 & 0.159 & 1734\\
K & -1.134 $\pm$ 0.018 & -2.499 $\pm$ 0.026 & 0.005 $\pm$ 0.011 & 0.163 & 1734\\
L & -1.239 $\pm$ 0.017 & -2.574 $\pm$ 0.025 & 0.006 $\pm$ 0.01 & 0.157 & 1734\\
L' & -1.242 $\pm$ 0.017 & -2.576 $\pm$ 0.025 & 0.007 $\pm$ 0.01 & 0.157 & 1734\\
M & -1.482 $\pm$ 0.015 & -2.842 $\pm$ 0.022 & 0.018 $\pm$ 0.009 & 0.14 & 1734\\
Bolometric & -0.176 $\pm$ 0.028 & -1.806 $\pm$ 0.041 & 0.022 $\pm$ 0.017 & 0.253 & 1734\\
\hline
\multicolumn{6}{c}{Convection set B}\\
\hline
U & 0.344 $\pm$ 0.035 & -0.958 $\pm$ 0.049 & 0.117 $\pm$ 0.022 & 0.271 & 1217\\
B & 0.245 $\pm$ 0.037 & -1.006 $\pm$ 0.051 & 0.035 $\pm$ 0.023 & 0.286 & 1217\\
V & -0.14 $\pm$ 0.032 & -1.372 $\pm$ 0.044 & -0.005 $\pm$ 0.02 & 0.246 & 1217\\
R & -0.375 $\pm$ 0.028 & -1.616 $\pm$ 0.039 & -0.01 $\pm$ 0.017 & 0.22 & 1217\\
I & -0.605 $\pm$ 0.026 & -1.833 $\pm$ 0.035 & -0.005 $\pm$ 0.016 & 0.198 & 1217\\
J & -0.935 $\pm$ 0.023 & -2.09 $\pm$ 0.031 & 0.002 $\pm$ 0.014 & 0.174 & 1217\\
H & -1.195 $\pm$ 0.02 & -2.386 $\pm$ 0.027 & 0.009 $\pm$ 0.012 & 0.153 & 1217\\
K & -1.148 $\pm$ 0.02 & -2.338 $\pm$ 0.028 & 0.009 $\pm$ 0.012 & 0.156 & 1217\\
L & -1.255 $\pm$ 0.02 & -2.419 $\pm$ 0.027 & 0.01 $\pm$ 0.012 & 0.152 & 1217\\
L' & -1.258 $\pm$ 0.02 & -2.421 $\pm$ 0.027 & 0.011 $\pm$ 0.012 & 0.152 & 1217\\
M & -1.508 $\pm$ 0.018 & -2.709 $\pm$ 0.025 & 0.022 $\pm$ 0.011 & 0.14 & 1217\\
Bolometric & -0.162 $\pm$ 0.029 & -1.58 $\pm$ 0.04 & 0.027 $\pm$ 0.018 & 0.222 & 1217\\
\hline
\multicolumn{6}{c}{Convection set C}\\
\hline
U & 0.414 $\pm$ 0.041 & -0.792 $\pm$ 0.056 & 0.129 $\pm$ 0.025 & 0.318 & 1292\\
B & 0.327 $\pm$ 0.041 & -0.935 $\pm$ 0.057 & 0.039 $\pm$ 0.025 & 0.323 & 1292\\
V & -0.077 $\pm$ 0.035 & -1.362 $\pm$ 0.048 & -0.008 $\pm$ 0.021 & 0.273 & 1292\\
R & -0.328 $\pm$ 0.031 & -1.618 $\pm$ 0.043 & -0.014 $\pm$ 0.019 & 0.242 & 1292\\
I & -0.57 $\pm$ 0.028 & -1.843 $\pm$ 0.038 & -0.009 $\pm$ 0.017 & 0.217 & 1292\\
J & -0.91 $\pm$ 0.024 & -2.142 $\pm$ 0.033 & -0.005 $\pm$ 0.015 & 0.188 & 1292\\
H & -1.184 $\pm$ 0.02 & -2.472 $\pm$ 0.028 & 0.002 $\pm$ 0.012 & 0.16 & 1292\\
K & -1.136 $\pm$ 0.021 & -2.415 $\pm$ 0.029 & 0.002 $\pm$ 0.013 & 0.164 & 1292\\
L & -1.246 $\pm$ 0.02 & -2.506 $\pm$ 0.028 & 0.003 $\pm$ 0.012 & 0.158 & 1292\\
L' & -1.248 $\pm$ 0.02 & -2.509 $\pm$ 0.028 & 0.004 $\pm$ 0.012 & 0.157 & 1292\\
M & -1.507 $\pm$ 0.018 & -2.823 $\pm$ 0.025 & 0.017 $\pm$ 0.011 & 0.14 & 1292\\
Bolometric & -0.111 $\pm$ 0.031 & -1.606 $\pm$ 0.043 & 0.024 $\pm$ 0.019 & 0.245 & 1292\\
\hline
\multicolumn{6}{c}{Convection set D}\\
\hline
U & 0.429 $\pm$ 0.038 & -0.762 $\pm$ 0.05 & 0.127 $\pm$ 0.024 & 0.28 & 1094\\
B & 0.334 $\pm$ 0.04 & -0.87 $\pm$ 0.052 & 0.033 $\pm$ 0.025 & 0.29 & 1094\\
V & -0.075 $\pm$ 0.034 & -1.28 $\pm$ 0.045 & -0.01 $\pm$ 0.021 & 0.251 & 1094\\
R & -0.327 $\pm$ 0.031 & -1.536 $\pm$ 0.04 & -0.015 $\pm$ 0.019 & 0.226 & 1094\\
I & -0.569 $\pm$ 0.028 & -1.76 $\pm$ 0.037 & -0.008 $\pm$ 0.017 & 0.205 & 1094\\
J & -0.912 $\pm$ 0.025 & -2.047 $\pm$ 0.033 & -0.001 $\pm$ 0.016 & 0.183 & 1094\\
H & -1.188 $\pm$ 0.022 & -2.369 $\pm$ 0.029 & 0.008 $\pm$ 0.014 & 0.161 & 1094\\
K & -1.139 $\pm$ 0.023 & -2.314 $\pm$ 0.029 & 0.008 $\pm$ 0.014 & 0.164 & 1094\\
L & -1.25 $\pm$ 0.022 & -2.402 $\pm$ 0.029 & 0.01 $\pm$ 0.014 & 0.16 & 1094\\
L' & -1.252 $\pm$ 0.022 & -2.405 $\pm$ 0.029 & 0.011 $\pm$ 0.014 & 0.16 & 1094\\
M & -1.513 $\pm$ 0.02 & -2.715 $\pm$ 0.026 & 0.026 $\pm$ 0.012 & 0.146 & 1094\\
Bolometric & -0.109 $\pm$ 0.031 & -1.511 $\pm$ 0.041 & 0.023 $\pm$ 0.019 & 0.227 & 1094\\
\hline
\end{tabular}}
\label{tab:B1}
\end{table}

\begin{table}
\caption{$PLZ$ relations of the mathematical form $M_\lambda=a+b\log(P)+c\mathrm{[Fe/H]}$ for BL~Her models in the high-metallicity regime ($Z=0.00424, 0.00834, 0.01300$) for different wavelengths using different convective parameter sets.}
\centering
\scalebox{0.85}{
\begin{tabular}{c c c c c c}
\hline\hline
Band & $a$ & $b$ & $c$ & $\sigma$ & $N$\\
\hline \hline
\multicolumn{6}{c}{Convection set A}\\
\hline
U & 0.432 $\pm$ 0.026 & -0.816 $\pm$ 0.062 & 0.442 $\pm$ 0.048 & 0.382 & 1532\\
B & 0.288 $\pm$ 0.024 & -1.187 $\pm$ 0.057 & 0.173 $\pm$ 0.045 & 0.353 & 1532\\
V & -0.118 $\pm$ 0.019 & -1.642 $\pm$ 0.046 & 0.043 $\pm$ 0.036 & 0.282 & 1532\\
R & -0.358 $\pm$ 0.017 & -1.873 $\pm$ 0.04 & 0.015 $\pm$ 0.031 & 0.245 & 1532\\
I & -0.579 $\pm$ 0.015 & -2.066 $\pm$ 0.035 & 0.027 $\pm$ 0.027 & 0.217 & 1532\\
J & -0.89 $\pm$ 0.012 & -2.347 $\pm$ 0.03 & 0.025 $\pm$ 0.023 & 0.182 & 1532\\
H & -1.135 $\pm$ 0.01 & -2.621 $\pm$ 0.025 & 0.039 $\pm$ 0.019 & 0.153 & 1532\\
K & -1.086 $\pm$ 0.011 & -2.583 $\pm$ 0.025 & 0.041 $\pm$ 0.02 & 0.157 & 1532\\
L & -1.192 $\pm$ 0.01 & -2.652 $\pm$ 0.025 & 0.044 $\pm$ 0.019 & 0.152 & 1532\\
L' & -1.193 $\pm$ 0.01 & -2.652 $\pm$ 0.025 & 0.048 $\pm$ 0.019 & 0.152 & 1532\\
M & -1.427 $\pm$ 0.01 & -2.833 $\pm$ 0.023 & 0.103 $\pm$ 0.018 & 0.14 & 1532\\
Bolometric & -0.108 $\pm$ 0.017 & -1.864 $\pm$ 0.04 & 0.092 $\pm$ 0.031 & 0.248 & 1532\\
\hline
\multicolumn{6}{c}{Convection set B}\\
\hline
U & 0.411 $\pm$ 0.027 & -0.428 $\pm$ 0.069 & 0.385 $\pm$ 0.053 & 0.35 & 1043\\
B & 0.269 $\pm$ 0.025 & -0.847 $\pm$ 0.064 & 0.1 $\pm$ 0.05 & 0.328 & 1043\\
V & -0.141 $\pm$ 0.02 & -1.372 $\pm$ 0.052 & -0.02 $\pm$ 0.04 & 0.267 & 1043\\
R & -0.383 $\pm$ 0.018 & -1.638 $\pm$ 0.046 & -0.041 $\pm$ 0.036 & 0.235 & 1043\\
I & -0.606 $\pm$ 0.016 & -1.856 $\pm$ 0.041 & -0.021 $\pm$ 0.032 & 0.21 & 1043\\
J & -0.921 $\pm$ 0.014 & -2.175 $\pm$ 0.035 & -0.02 $\pm$ 0.027 & 0.18 & 1043\\
H & -1.168 $\pm$ 0.012 & -2.488 $\pm$ 0.03 & 0.002 $\pm$ 0.024 & 0.155 & 1043\\
K & -1.119 $\pm$ 0.012 & -2.445 $\pm$ 0.031 & 0.002 $\pm$ 0.024 & 0.158 & 1043\\
L & -1.225 $\pm$ 0.012 & -2.522 $\pm$ 0.03 & 0.008 $\pm$ 0.023 & 0.153 & 1043\\
L' & -1.227 $\pm$ 0.012 & -2.522 $\pm$ 0.03 & 0.011 $\pm$ 0.023 & 0.154 & 1043\\
M & -1.462 $\pm$ 0.011 & -2.727 $\pm$ 0.028 & 0.076 $\pm$ 0.022 & 0.143 & 1043\\
Bolometric & -0.134 $\pm$ 0.018 & -1.622 $\pm$ 0.046 & 0.032 $\pm$ 0.036 & 0.237 & 1043\\
\hline
\multicolumn{6}{c}{Convection set C}\\
\hline
U & 0.605 $\pm$ 0.031 & -0.553 $\pm$ 0.074 & 0.46 $\pm$ 0.06 & 0.438 & 1340\\
B & 0.465 $\pm$ 0.027 & -1.097 $\pm$ 0.064 & 0.143 $\pm$ 0.052 & 0.38 & 1340\\
V & 0.012 $\pm$ 0.021 & -1.602 $\pm$ 0.05 & 0.009 $\pm$ 0.041 & 0.3 & 1340\\
R & -0.261 $\pm$ 0.019 & -1.828 $\pm$ 0.044 & -0.016 $\pm$ 0.036 & 0.263 & 1340\\
I & -0.509 $\pm$ 0.017 & -2.008 $\pm$ 0.039 & 0.003 $\pm$ 0.032 & 0.234 & 1340\\
J & -0.849 $\pm$ 0.014 & -2.324 $\pm$ 0.032 & -0.003 $\pm$ 0.026 & 0.192 & 1340\\
H & -1.127 $\pm$ 0.011 & -2.601 $\pm$ 0.027 & 0.02 $\pm$ 0.022 & 0.16 & 1340\\
K & -1.074 $\pm$ 0.012 & -2.563 $\pm$ 0.027 & 0.019 $\pm$ 0.022 & 0.163 & 1340\\
L & -1.186 $\pm$ 0.011 & -2.634 $\pm$ 0.026 & 0.024 $\pm$ 0.021 & 0.158 & 1340\\
L' & -1.187 $\pm$ 0.011 & -2.635 $\pm$ 0.027 & 0.028 $\pm$ 0.021 & 0.158 & 1340\\
M & -1.447 $\pm$ 0.01 & -2.796 $\pm$ 0.024 & 0.101 $\pm$ 0.02 & 0.145 & 1340\\
Bolometric & -0.004 $\pm$ 0.019 & -1.843 $\pm$ 0.044 & 0.06 $\pm$ 0.036 & 0.262 & 1340\\
\hline
\multicolumn{6}{c}{Convection set D}\\
\hline
U & 0.545 $\pm$ 0.034 & -0.275 $\pm$ 0.083 & 0.399 $\pm$ 0.066 & 0.428 & 1028\\
B & 0.412 $\pm$ 0.03 & -0.844 $\pm$ 0.073 & 0.078 $\pm$ 0.058 & 0.375 & 1028\\
V & -0.028 $\pm$ 0.023 & -1.403 $\pm$ 0.058 & -0.043 $\pm$ 0.046 & 0.297 & 1028\\
R & -0.294 $\pm$ 0.02 & -1.659 $\pm$ 0.05 & -0.06 $\pm$ 0.04 & 0.26 & 1028\\
I & -0.536 $\pm$ 0.018 & -1.862 $\pm$ 0.045 & -0.034 $\pm$ 0.036 & 0.231 & 1028\\
J & -0.87 $\pm$ 0.015 & -2.208 $\pm$ 0.037 & -0.034 $\pm$ 0.03 & 0.191 & 1028\\
H & -1.14 $\pm$ 0.013 & -2.519 $\pm$ 0.031 & -0.003 $\pm$ 0.025 & 0.16 & 1028\\
K & -1.089 $\pm$ 0.013 & -2.477 $\pm$ 0.032 & -0.005 $\pm$ 0.025 & 0.164 & 1028\\
L & -1.199 $\pm$ 0.012 & -2.554 $\pm$ 0.031 & 0.003 $\pm$ 0.025 & 0.158 & 1028\\
L' & -1.2 $\pm$ 0.012 & -2.555 $\pm$ 0.031 & 0.006 $\pm$ 0.025 & 0.158 & 1028\\
M & -1.454 $\pm$ 0.012 & -2.739 $\pm$ 0.028 & 0.088 $\pm$ 0.023 & 0.147 & 1028\\
Bolometric & -0.038 $\pm$ 0.02 & -1.667 $\pm$ 0.05 & 0.011 $\pm$ 0.04 & 0.26 & 1028\\
\hline
\end{tabular}}
\label{tab:B2}
\end{table}

\label{lastpage}
\end{document}